\documentclass{aa}  

\usepackage{graphicx}
\usepackage{txfonts}
\usepackage[colorlinks=True,pdfborder={0 0 0}, linkcolor=blue,citecolor=blue]{hyperref}

\def\tvi(#1,#2){\vrule height #1pt depth #2pt width 0pt}

\def\a{a}
\def\b{b}
\def\v2{v_{\rm sh}}

\def\M{{\cal M}}

\def\N{N}

\begin{document}

   \title{Analytic insight into the physics of SASI}

   \subtitle{I. Shock instability in a non-rotating stellar core}

   \author{T. Foglizzo
          }

   \institute{Universit\'e Paris-Saclay, Universit\'e Paris Cit\'e, CEA, CNRS, AIM, 91191, Gif-sur-Yvette, France\\
              \email{foglizzo@cea.fr}
             }

   \date{Received 18 September 2024; accepted 23 October 2024}

 
  \abstract
   {During the core collapse of a massive star just before its supernova explosion, the amplification of asymmetric motions by the standing accretion shock instability (SASI) imprints on the neutrino flux and the gravitational waves a frequency signature carrying direct information on the explosion process.  }
   {The physical interpretation of this multi-messenger signature requires a detailed understanding of the instability mechanism. }
   {A perturbative analysis is used to characterize the properties of SASI, and assess the effect of the region of neutronization above the surface of the proto-neutron star. The eigenfrequencies of the most unstable modes are compared to those obtained in an adiabatic approximation where neutrino interactions are neglected above the neutrinosphere. The differential system is solved analytically using a Wronskian method and approximated asymptotically for a large shock radius.}
   {The oscillation period of SASI is well fitted with a simple analytic function of the shock radius, the radius of maximum deceleration and the mass of the proto-neutron star. The oscillation period is weakly dependent on the parametrized cooling function which however affects the SASI growth rate. The general properties of SASI eigenmodes are described using an adiabatic model. In this approximation the eigenvalue problem is formulated as a self-forced oscillator. The forcing agent is the radial advection of  baroclinic vorticity perturbations and entropy perturbations produced by the shock oscillation. The differential system defining the eigenfrequencies is reduced to a single integral equation. Its analytical approximation sheds light on the radially extended character of the region of advective-acoustic coupling. The simplicity of this adiabatic formalism opens new perspectives to investigate the effect of stellar rotation and non-adiabatic processes on SASI. }
   {}


   \maketitle
%

\section{Introduction}

  The explosive death of massive stars is sensitive to the development of hydrodynamical instabilities which break the spherical symmetry of the stellar core, affect the efficiency of neutrino energy absorption and generate turbulence \citep{Muller2020,Janka2016,Burrows2021}. A spherically symmetric scenario based on radial motions seems possible only for the lightest progenitors \citep{Kitaura2006,Stockinger2020}.  Asymmetric motions contribute to the kick and spin of the neutron star \citep{Muller_Tauris2019, Janka2017}, the emission of gravitational waves \citep{Kotake2017} and a modulation of neutrino emission \citep{Muller2019a,Tamborra2019}. The correlation of instability signatures in the gravitational waves and neutrino signals has been considered by \cite{Drago2023} to improve the detection efficiency of gravitational waves from nearby supernovae. Ultimately, the information encoded in the gravitational waves and neutrino signals can be used to recover the properties of the dying star and its explosion mechanism \citep{Powell2022}. 
The computational cost of 3D numerical simulations precludes a systematic coverage of the large parameter space describing the initial conditions of the stellar core-collapse. Understanding the underlying mechanism of instabilities is essential to extrapolate the results of sparse numerical simulation, evaluate the impact of additional physical ingredients, and design effective prescriptions for parametric studies  \citep{Muller2019b}.\\
Among the hydrodynamical instabilities at work during the phase of stalled accretion shock, the Standing Accretion Shock Instability \citep{Blondin2003} is able to introduce coherent transverse motions with a large angular scale, growing over a timescale related to the advection time from the shock to the neutron star surface. The mechanism of SASI has been described as an advective-acoustic cycle between the shock and the vicinity of the proto-neutron star \citep{Foglizzo2007, Fernandez2009a,Scheck2008}. The analytical understanding of SASI eigenfrequencies in the WKB approximation is however restricted to the limit of a large shock radius for high frequency overtones \citep{Foglizzo2007} while the lowest frequency fundamental mode is often the most unstable. A fully analytic solution including the fundamental mode was obtained only in a very idealized model where the stationary flow is plane parallel and uniform except in a compact region of deceleration mimicking the vicinity of the neutron star \citep{Foglizzo2009} (hereafter F09). This toy model neglected the flow gradients that are actually extended all the way from the shock to the neutron star and the non-adiabatic character of the neutrino processes taking place in the vicinity of the neutron star. Despite these limitations, this model illustrated the interplay of the advective-acoutic and the purely acoustic cycles, and the expected phase mixing taking place at high frequency. It was used by \cite{Guilet2012} to interpret the frequency spacing of the eigenspectrum in the framework of the advective-acoustic mechanism rather than a purely acoustic process. The physical understanding of SASI led \cite{Muller_Janka2014} to define an empirical formula for the SASI oscillation period, in order to interpret the modulation of the neutrino signal. Directly proportional to an approximation of the advection time from the shock to the neutron star surface, it was tested on a 2D numerical simulation of the collapse of a $25M_{\odot}$ progenitor. However its generality for other progenitors, proposed by \cite{Muller2019a}, has not been established yet.

The present study aims at improving our understanding of the fundamental mode of SASI in spherical geometry in order to better identify the physical parameters governing its oscillation frequency. Simple cooling functions mimicking neutrino emission will be used to assess the role of the cooling region. An adiabatic approximation will also be used to assess the role of the advective-acoustic coupling in the radially extended region between the shock and the proto-neutron star. The adiabatic approximation is motivated by the adiabatic simulations of \cite{Blondin_Mezzacappa2007} and by the shallow water experiment \citep{Foglizzo2012, Foglizzo2015}, which both suggest that several properties of SASI may be understood using adiabatic equations. The adiabatic approximation has also been used to analyze the impact of general relativistic corrections on SASI eigenfrequencies \citep{Dunham2024}.

The set of differential equations defining the eigenfrequencies of spherical accretion is recalled in Sect.~2 with particular attention to the radial extension of the non-adiabatic cooling layer and its impact on the oscillation period of SASI. The eigenfrequencies calculated in the adiabatic approximation are compared to those including neutrino losses in Sect.~3. The adiabatic model is formulated as a self-forced oscillator in Sect.~4, with eigenfrequencies defined by an integral equation. An analytic approximation of this equation is obtained and analyzed in Sect.~5. Conclusions and perspectives are formulated in Sect.~6.

\section{Stationary accretion with non-adiabatic cooling}

\subsection{Stationary flow}

We use the same general framework as \cite{Blondin2003, Foglizzo2007,Blondin2017} to 
describe the phase where the accretion shock stalls at the radius $r_{\rm sh}$. Neutrino absorption is neglected and neutrino emission near the proto-neutron star radius $r_{\rm ns}$ is idealized with a cooling function of the density $\rho$ and pressure $p$: 
\begin{eqnarray}
{\cal L}= -A_{\rm cool}\rho^{\beta-\alpha} p^\alpha. \label{def_cooling}
\end{eqnarray}
The collapsing stellar core immediately after bounce is modelled as a perfect gas with an adiabatic index $\gamma=4/3$, dominated by the degeneracy pressure of relativistic electrons. The gravitational potential $\Phi\equiv -GM_{\rm ns}/r$ is assumed to be dominated by the mass $M_{\rm ns}$ of the proto-neutron star in the Newtonian approximation for simplicity. The self-gravity of the accreting gas and the increase of $M_{\rm ns}$ with time are neglected. 
The dimensionless measure of the entropy $S$ is defined by:
\begin{eqnarray}
S&\equiv& {1\over\gamma-1}\log{p\over\rho^\gamma}\label{def_entropy}.
\end{eqnarray}
The stationary flow is described by the mass conservation, the entropy equation and the Euler equation:
\begin{eqnarray}
     \rho v &=& \left({r_{\rm sh}\over r}\right)^g\rho_{\rm sh} v_{\rm sh}\ , \label{eq_mass_stat}\\ 
    {\partial S\over\partial r} &=& {{\cal L}\over p v}\ , \label{eq_entropy_stat}\\
    {\partial\over\partial r} \left( {v^2\over 2}  + {c^2\over\gamma - 1} + \Phi \right) &=& {{\cal L}\over\rho v}\ ,\label{eq_euler_stat}
\end{eqnarray}
where $c\equiv (\gamma p/\rho)^{1/2}$ is the sound speed and $v<0$ the radial velocity. The geometrical parameter $g=2$ accounting for the spherical geometry is noted as a parameter in this section and Sect.~\ref{sect_adiabatic} to keep track of its impact on the equations in the spirit of \cite{Walk2023}. The subscript "sh" in Eq.~(\ref{eq_mass_stat}) refers to quantities immediately below the shock, and the subscript "1" refers to pre-shock quantities.
The jump conditions at the shock follow from the conservation of mass flux and momentum flux, taking into account the energy lost across the shock through nuclear dissociation. This latter is modeled as in \cite{Fernandez2009a,Fernandez2014} by the parameter $\varepsilon$ measuring the energy loss $\Delta e_{\rm disso}$ per unit of mass in units of the pre-shock kinetic energy density $v_1^2/2$ as in \cite{Huete2018}: 
\begin{eqnarray}
\varepsilon&\equiv& {\Delta e_{\rm disso}\over v_1^2/2}.\label{norm_disso}
\end{eqnarray}
The effect of $\varepsilon$ on the post-shock Mach number $\M_{\rm sh}$ and the Rankine-Hugoniot conditions is recalled in Appendix~\ref{sec:Adiff}, following Eqs.~(A4-A6) of \cite{Foglizzo2006}.
The pre-shock deceleration effect of pressure is neglected, with $v_1^2\sim 2GM_{\rm ns}/r_{\rm sh}$.
Defining the Mach number $\M\equiv |v|/c$ as positive, we assume a strong adiabatic shock $\M_1\gg1$ in the numerical calculations of this section.\\
\begin{figure}
\centering
\includegraphics[width=\columnwidth]{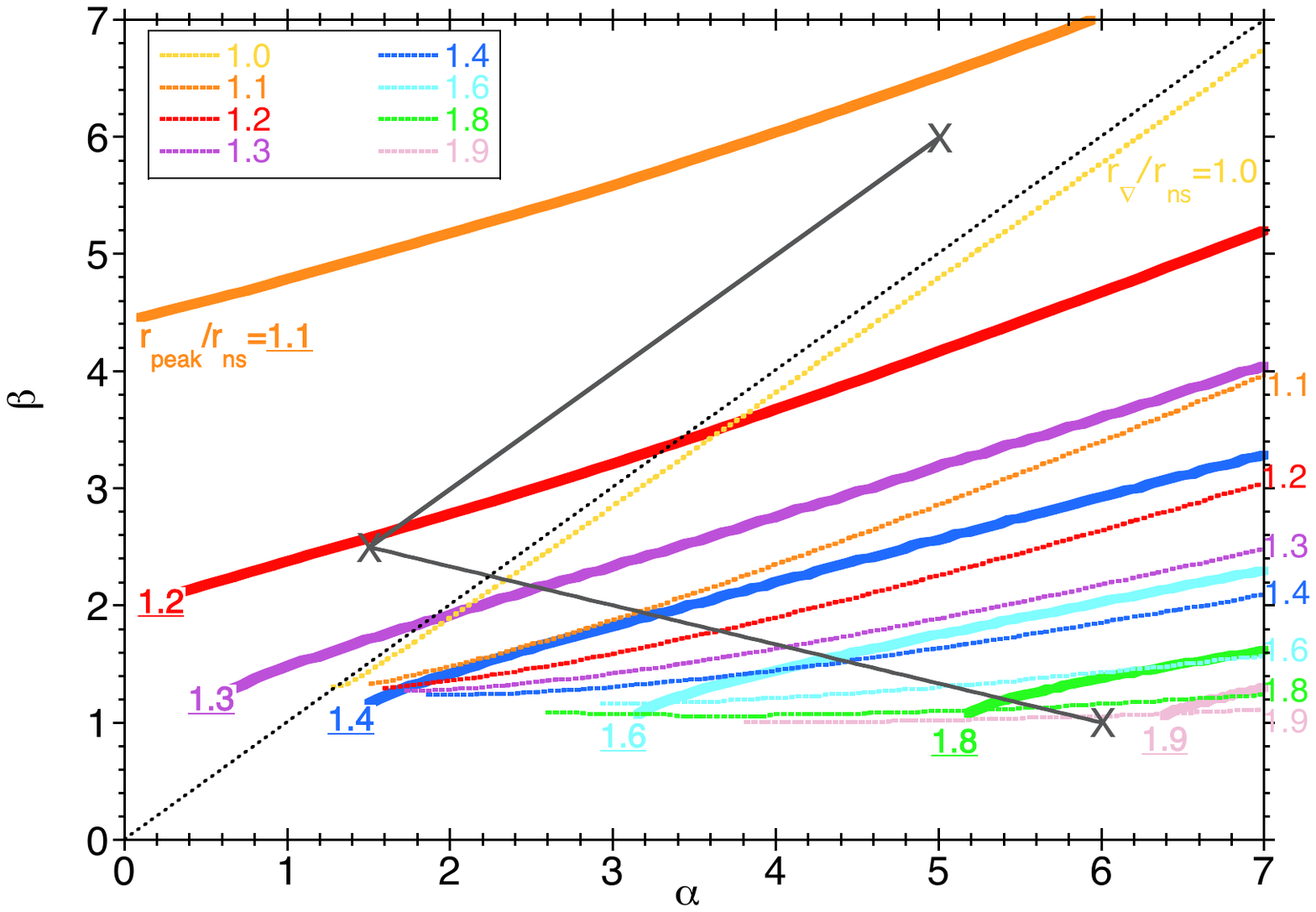}
\caption{Ratios $r_{\rm peak}/r_{\rm ns}$ (thick solid lines) and $r_{\nabla}/r_{\rm ns}$ (thin dotted lines) depending on the coefficients $(\alpha,\beta)$ which define the cooling function in Eq.~(\ref{def_cooling}). The contour lines are calculated for $\gamma=4/3$ in spherical geometry, with $r_{\rm sh}/r_{\rm ns}=10$, without dissociation ($\varepsilon=0$). The values of $r_{\rm peak}/r_{\rm ns}$ and $r_{\nabla}/r_{\rm ns}$ are indicated with the same colour code. The cooling parameters $(\alpha,\beta)=(3/2,5/2)$, $(6,1)$ and $(5,6)$ used in Fig.~\ref{fig_w_alpha} are indicated with crosses for reference, connected by grey lines. The black dotted line marks the threshold $\beta=\alpha$ above which the advection time from $r_{\rm peak}$ to $r_{\rm ns}$ is finite and $r_\nabla=r_{\rm ns}$. }
\label{fig_profile_cooling}
\end{figure}
The adiabatic compression of the post-shock flow by the gravitational potential $\Phi$ produces an inward increase of the enthalpy $c^2/(\gamma-1)$ and thus an increase of the gas temperature $T\propto c^2$ according to the Bernoulli equation (\ref{eq_euler_stat}). In our model of stationary accretion the temperature profile displays a maximum at a radius noted $r_{\rm peak}$ where the energy losses by neutrino emission balance the adiabatic heating due to the gravitational compression. 
The width of the cooling layer depends a priori on the parameters $(\alpha,\beta)$ of the cooling function ${\cal L}$, the mass accretion rate and the geometry. 
With $\alpha=3/2$ and $\beta=5/2$, \cite{Walk2023} noted that both the radius $r_{\rm peak}\sim 1.2r_{\rm ns}$ of maximum temperature and the typical width $\Delta r_{\rm peak}\sim 1.5r_{\rm ns}$ of the temperature profile measured at half maximum seemed surprisingly independent of both the mass accretion rate and the shock radius, and also seemed independent of the geometry. A closer inspection confirms that the location of $r_{\rm peak}$ varies by less than $1\%$ between the cylindrical and spherical geometries, and varies by less than $0.5\%$ with $r_{\rm sh}$ for $r_{\rm sh}/r_{\rm ns}>3$. 

Figure~\ref{fig_profile_cooling} compares $r_{\rm peak}$ to the radius of maximum deceleration noted $r_\nabla$, with $r_\nabla<r_{\rm peak}$ over a large range of parameters $(\alpha,\beta)$ except for $\beta$ approaching unity. Noting that $r_\nabla=r_{\rm ns}$ for $\alpha<\beta$ \citep{Foglizzo2007}, Fig.~\ref{fig_profile_cooling} indicates that $r_{\rm peak}/r_\nabla\sim 1.2$ for $(\alpha,\beta)=(3/2,5/2)$, and $r_{\rm peak}/r_\nabla\sim 1.0$ for $(\alpha,\beta)=(6,1)$. 
In Sect.~\ref{Sect_perturb_cooling} the parameters $(\alpha,\beta)$ are varied continuously from $(3/2,5/2)$ to $(6,1)$ and from $(3/2,5/2)$ to $(5,6)$ in order to evaluate their impact on SASI properties.

\subsection{Perturbed flow}
\label{Sect_perturb_cooling}

The perturbations of the stationary flow are characterized by the wavenumbers $\ell,m$ of spherical harmonics and a complex eigenfrequency $\omega$. The eigenfrequency is independent of the azimuthal wavenumber $|m|\le \ell$ as established in \cite{Foglizzo2007}. We use the same physical variables as in \cite{Foglizzo2006} guided by analytical simplicity: the perturbation of the Bernoulli constant $\delta f$, the perturbed mass flux $\delta h$, the perturbed dimensionless entropy $\delta S$. The perturbation $\delta K$ is a combination of perturbed entropy $\delta S$ and the quantity $\delta w_\perp$ defined from the radial component of the curl of the vorticity perturbation $\delta w\equiv\nabla\times\delta v$:
\begin{eqnarray}
\delta f &\equiv& v\delta v_r + {\delta c^2\over\gamma-1} \ , \label{defdf0}\\
\delta h &\equiv& {\delta v_r\over v} + {\delta \rho\over\rho} \ , \label{defdh0}\\
\delta S &\equiv& {1\over\gamma-1} {\delta c^2\over c^2} - {\delta \rho\over\rho}\ , \label{defdS0}\\
\delta w_\perp&\equiv& r(\nabla\times\delta w)_r,
\label{def_dwperp}\\
&=&
{1\over \sin\theta}
\left\lbrack{\partial\over\partial\theta}(\delta w_\varphi\sin\theta)-im  \delta w_\theta\right\rbrack
,
\\
{\delta K} &\equiv& rv\delta w_\perp
+  \ell(\ell+1){c^2\over\gamma}\delta S \ ,\label{defdK0}
\end{eqnarray}
We introduce the quantity $\delta A$ defined with the divergence of the ortho-radial velocity $\delta v_\perp\equiv(0,\delta v_\theta,\delta v_\varphi)$
\begin{eqnarray}
\delta A
&\equiv& r^2\nabla\cdot\delta v_\perp,\\
&=& {r\over \sin\theta}
\left\lbrack{\partial\over\partial\theta}(\delta v_\theta\sin\theta)+im  \delta v_\varphi\right\rbrack.\label{def_dA}
\end{eqnarray} 
We establish in Appendix~\ref{append_nonax} that $\delta A$ and $\delta w_\perp$ are related to the perturbation $\delta v_r$ of the radial velocity as follows: 
\begin{eqnarray}
\delta v_r&=&
{r\delta w_\perp\over \ell(\ell+1)}
-{1\over \ell(\ell+1)}{\partial \delta A\over\partial r}
 .\label{dvrw_main}
\end{eqnarray}
This equation invites us to interpret $\delta A/\ell(\ell+1)$ as the potential defining the compressible part of the perturbed velocity, and $r\delta w_\perp/\ell(\ell+1)$ as the rotational contribution to the radial velocity perturbation.
Using in Appendix~\ref{append_nonax} the transverse components of the Euler equation (Eqs.~\ref{ddfdtheta}-\ref{ddfdphi}) with Eq.~(\ref{defdK0}), we note that $\delta A$ is related to $\delta K$ and $\delta f$ by
\begin{eqnarray}
\delta K=i\omega\delta A+\ell(\ell+1)\delta f.
\label{AFK}
\end{eqnarray} 
The differential system satisfied by ($\delta A,\delta h,\delta S,\delta K$) is as follows:
\begin{eqnarray}
\left({1-\M^2\over v}{\partial \over\partial r}+{i\omega\over c^2}\right){\delta A\over \ell(\ell+1)}
=
-\delta h
\nonumber\\
-\left(\gamma-1+{1\over\M^2}\right){\delta S\over\gamma}
+{\delta K\over v^2\ell(\ell+1)} 
,\label{dAdr_main}\\
\left({1-\M^2\over v}{\partial \over\partial r}+{i\omega\over c^2}\right) \delta h
={\omega^2-\omega_{\rm Lamb}^2\over v^2c^2}
{\delta A\over \ell(\ell+1)}
\nonumber\\
-{i\omega\over v^2}\delta S
+{i\omega\over v^2c^2}
{\delta K\over \ell(\ell+1)}
\label{dhdr_main}
,\\
\left({\partial \over\partial r}-{i\omega\over v}\right)
\delta S
=
 \delta \left( {{\cal L}\over p v } \right),\label{dSdr_main} \\
\left({\partial \over\partial r}-{i\omega\over v}\right)
{\delta K\over\ell(\ell+1)}
=
\delta \left( {{\cal L}\over \rho v } \right).\label{dKdr_main}
\end{eqnarray} 
It includes explicit non-adiabatic terms only in the Eqs.~(\ref{dSdr_main}-\ref{dKdr_main}) governing $\delta S$ and $\delta K$. In Eq.~(\ref{dhdr_main}) the Lamb frequency $\omega_{\rm Lamb}$ associated to the spherical harmonic of order $\ell$ defines the turning point of non-radial acoustic waves: 
\begin{eqnarray}
\omega_{\rm Lamb}^2&\equiv& \ell(\ell+1){c^2\over r^2}(1-\M^2).\label{defLamb}
\end{eqnarray} 
The boundary conditions at the shock are reformulated from Eqs.~(28,29,E7,E8) in \cite{Foglizzo2006} using Eq.~(\ref{AFK}):
\begin{eqnarray}
{\delta A_{\rm sh}\over\ell(\ell+1)} &=&   
-\left(1-{\v2\over v_{1}}\right)v_{1}\Delta \zeta 
\ ,\label{dAsh}\\
\delta h_{\rm sh} &=&   
\left(1-{\v2\over v_{1}}\right)
{\Delta v\over v_{\rm sh}}   
\ ,\label{dhsh}\\
\delta S_{\rm sh} &=& 
\gamma 
{v_1\over c_{\rm sh}^2}
(i\omega +\omega_\Phi)
\Delta\zeta
\left(1-{\v2\over v_{1}}\right)^2
,\label{dSsh_zeta}\\
\delta K_{\rm sh} &=& - \ell(\ell+1)\Delta \zeta {c^2_{\rm sh}\over\gamma}
\left\lbrack\nabla S\right\rbrack^{\rm sh}_{1}
\label{dKsh}
\ .
\end{eqnarray} 
$\Delta\zeta$ and $\Delta v\equiv -i\omega\Delta\zeta$ are the shock displacement and velocity.
The reference frequency $\omega_\Phi$ in Eq.~(\ref{dSsh_zeta}) is defined by:
\begin{eqnarray}
{\omega_\Phi r_{\rm sh}\over |v_{\rm sh}|}
\equiv
{1\over2}{v_1\over\v2}
{ {2r_{\rm sh}\over v_1^2}{{\rm d}\Phi\over {\rm d} r}
-4{\v2\over v_1}
\over
1-{\v2\over v_{1}}
}
+
{{\v2\over v_1}\over\gamma \M_{\rm sh}^2}
{
r_{\rm sh}\left\lbrack\nabla S\right\rbrack^{\rm sh}_{1}
\over
\left(1-{\v2\over v_{1}}\right)^2
}
.
\end{eqnarray}
$\omega_\Phi$ defines the threshold frequency separating the regime ($\omega\ll\omega_\Phi$) where the phase of $\delta S_{\rm sh}$ is opposed to $\Delta\zeta$, from the regime ($\omega\gg \omega_\Phi$) where it is set by $\Delta v$.\\
The first two differential equations (\ref{dAdr_main}-\ref{dhdr_main}) are transformed into the following second order differential equation:
\begin{eqnarray}
\left\lbrack\left(
{1-\M^2\over v}
{\partial\over\partial r}+{i\omega\over c^2}\right)^2
+
{\omega^2-\omega_{\rm Lamb}^2\over v^2c^2}\right\rbrack
\delta A
=
\nonumber\\
{\partial\over\partial r}
{r\delta w_\perp\over v}
-\ell(\ell+1)
{\gamma-1\over\gamma}
\delta \left( {{\cal L}\over p v} \right)
,\label{eqdiffAcool_main}
\end{eqnarray} 
We recognize on the left-hand-side of Eq.~(\ref{eqdiffAcool_main}) an acoustic oscillator modified by advection, with a forcing on the right-hand-side driven by vorticity perturbations and by the perturbation of neutrino cooling. Interestingly, part of this forcing can be studied analytically in the adiabatic approximation defined in Sect.~\ref{sect_adiabatic}.\\
The variables $\delta v_r$, $\delta \rho$, $\delta c$, $\delta p$ are deduced from $\delta A,\delta h,\delta S,\delta K$ using Eqs.~(\ref{dvr}-\ref{dP}) also recalled in Appendix~\ref{append_nonax}. In particular using Eq.~(\ref{dvr}) to express the lower boundary condition $\delta v_r(r_{\rm ns})=0$ leads to
\begin{eqnarray}
\left(c^2\delta h+c^2\delta S-\delta f \right)_{\rm ns}
=0,\label{inner_BC}
\end{eqnarray}
With a vanishing sound speed $c_{\rm ns}=0$ resulting from the hypothesis of stationary flow, the lower boundary condition $\delta v_r(r_{\rm ns})=0$ is equivalent to $\delta f_{\rm ns}= 0$.\\
The boundary value problem is solved numerically using a shooting method from the shock to the inner boundary, iterating over the value of the complex eigenfrequency $\omega$.
The mathematical singularity at $r_{\rm ns}$ is overcome numerically by using $\log \mathcal{M}$ as integration variable like in \cite{Foglizzo2007}. The inner boundary is then defined as the point where the Mach number has reached a sufficiently small value ($\mathcal{M} \sim 10^{-9}$). 
\\
\begin{figure}
\centering
\includegraphics[width=\columnwidth]{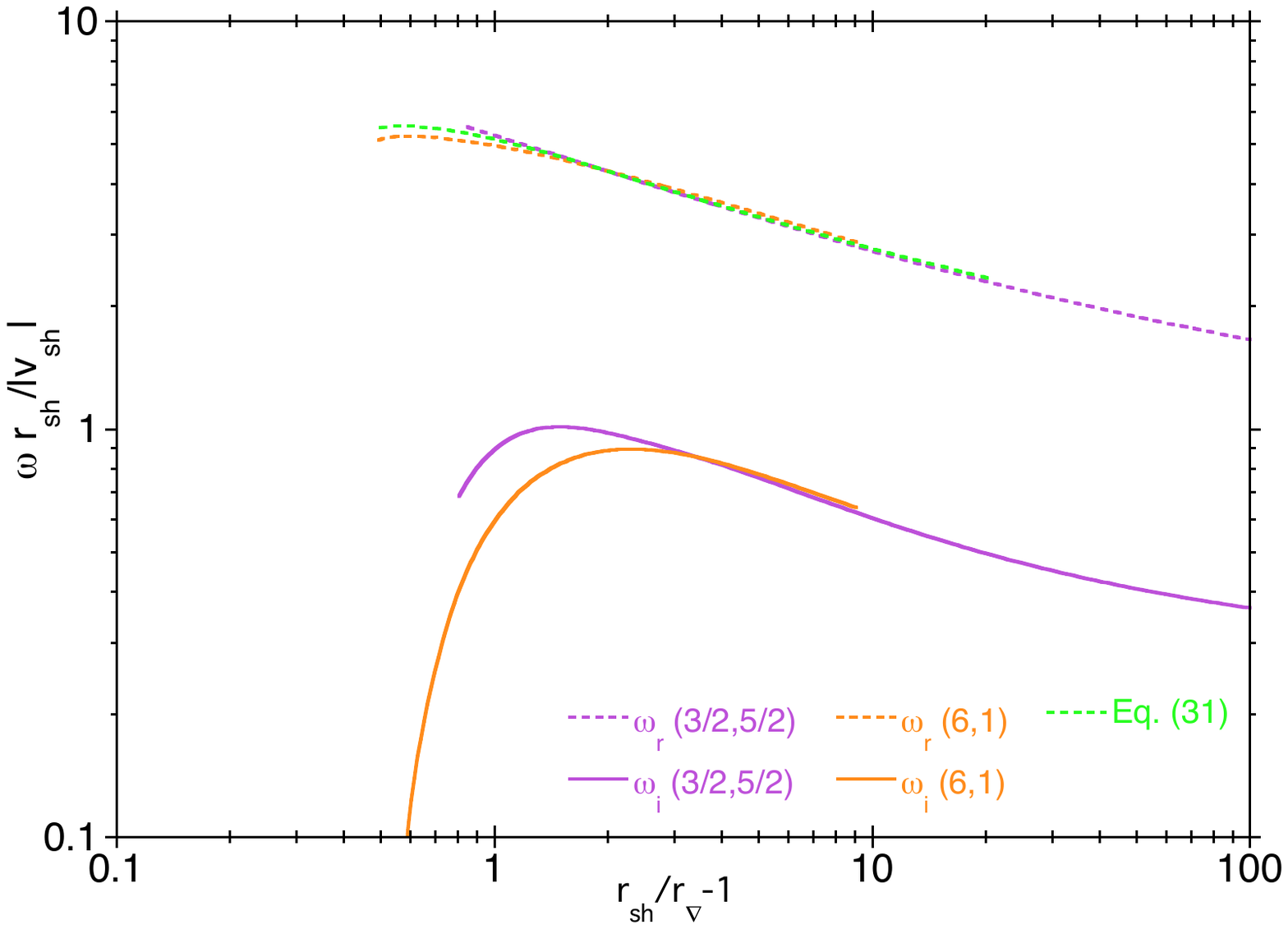}
\caption{Growth rate (solid lines) and oscillation frequency (dashed lines) of the fundamental SASI mode, normalized by $|v_{\rm sh}|/r_{\rm sh}$, calculated with $\varepsilon=0$ for the cooling parameters $(3/2,5/2)$ (purple) and $(6,1)$ (orange) and displayed as functions of $r_{\rm sh}/r_\nabla$. The analytic fitting formula (\ref{eqfitwr}) for $\omega_r$ is displayed with a green dashed line. }
\label{fig_fitwr}
\end{figure}
\begin{figure}
\centering
\includegraphics[width=\columnwidth]{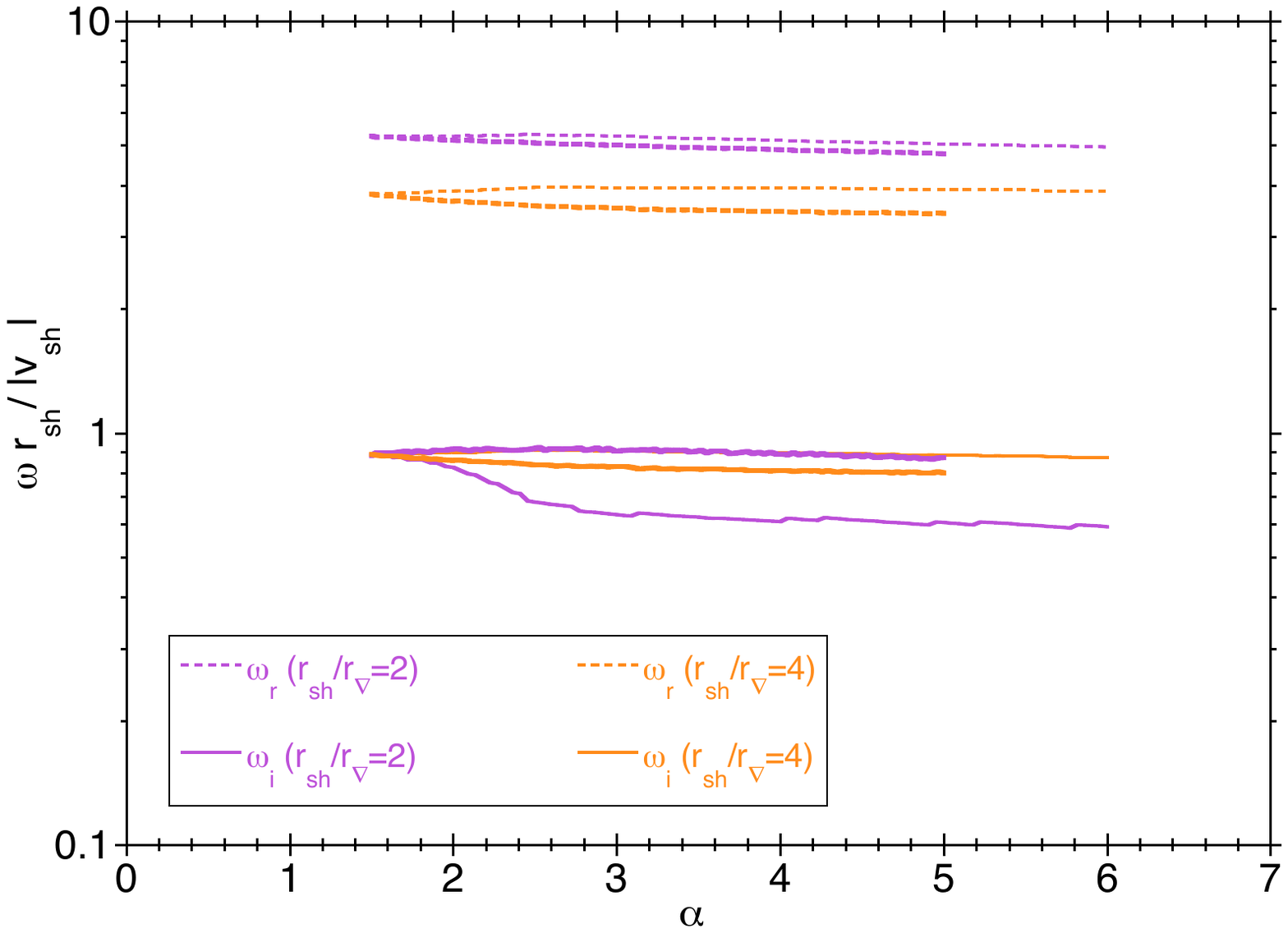}
\caption{Growth rate (solid lines) and oscillation frequency (dashed lines) of the fundamental SASI mode, normalized by $|v_{\rm sh}|/r_{\rm sh}$, calculated with $r_{\rm sh}/r_\nabla=2$ (purple lines) and $r_{\rm sh}/r_\nabla=4$ (orange lines) for a range of cooling parameters $(\alpha,\beta)$ varying continuously from $(3/2,5/2)$ to $(6,1)$ (thin lines) and from $(3/2,5/2)$ to $(5,6)$ (thick lines), as indicated in Fig.~\ref{fig_profile_cooling}. }
\label{fig_w_alpha}
\end{figure}
The analysis of \cite{Foglizzo2007} noted the similar properties of the SASI harmonics as functions of $r_{\rm sh}/r_\nabla$ with $(\alpha,\beta)=(3/2,5/2)$ and $(6,1)$. We extend this analysis for the fundamental mode and remark in Fig.~\ref{fig_fitwr} that the normalized growth rate and oscillation frequency are asymptotically independent of $(\alpha,\beta)$ for $r_{\rm sh}/r_\nabla\gg 1$, i.e. when the cooling layer is thin compared to the shock radius. This surprising property is checked by varying continuously $(\alpha,\beta)$ from $(3/2,5/2)$ to $(6,1)$ and from $(3/2,5/2)$ to $(5,6)$ in Fig.~\ref{fig_w_alpha}. It is remarkable that the normalized oscillation frequency seems independent of the cooling process down to a very low ratio $r_{\rm sh}/r_\nabla$: the following analytic formula $\omega_r^{\rm fit}$ provides an approximation for $\omega_r$ within $3\%$ for $1.8<r_{\rm sh}/r_\nabla<10$:
\begin{eqnarray}
\tau_{\rm adv}^{\rm adiab}\equiv  {r_{\rm sh}\over |v_{\rm sh}|}\log{r_{\rm sh}\over r_\nabla},\\
R_1\equiv\log\left({r_{\rm sh}\over r_\nabla}-1\right),\\
\omega_r^{\rm fit}\equiv {2\pi\over \tau_{\rm adv}^{\rm adiab}}
\left(
0.56773+0.28628R_1-0.031763R_1^2
\right).
\label{eqfitwr}
\end{eqnarray}
The adiabatic advection time $\tau_{\rm adv}^{\rm adiab}$ corresponds to a velocity profile increasing linearly with radius as expected asymptotically in spherical geometry for $\gamma=4/3$ (Eq.~19 in \cite{Walk2023}). 

The important role of $r_\nabla$ rather than $r_{\rm ns}$ was also noted in \cite{Scheck2008} where the strength of SASI in numerical simulations seemed correlated with the abruptness of the deceleration close to the proto-neutron star and confirmed in the toy model used by F09 and \cite{Guilet2012}. 

Figures~\ref{fig_fitwr} and \ref{fig_w_alpha} also show that the detailed cooling process does affect the growth rate of SASI (solid lines) more than its oscillation frequency (dashed lines) when $r_{\rm sh}<4r_\nabla$. \\
\begin{figure}
\centering
\includegraphics[width=\columnwidth]{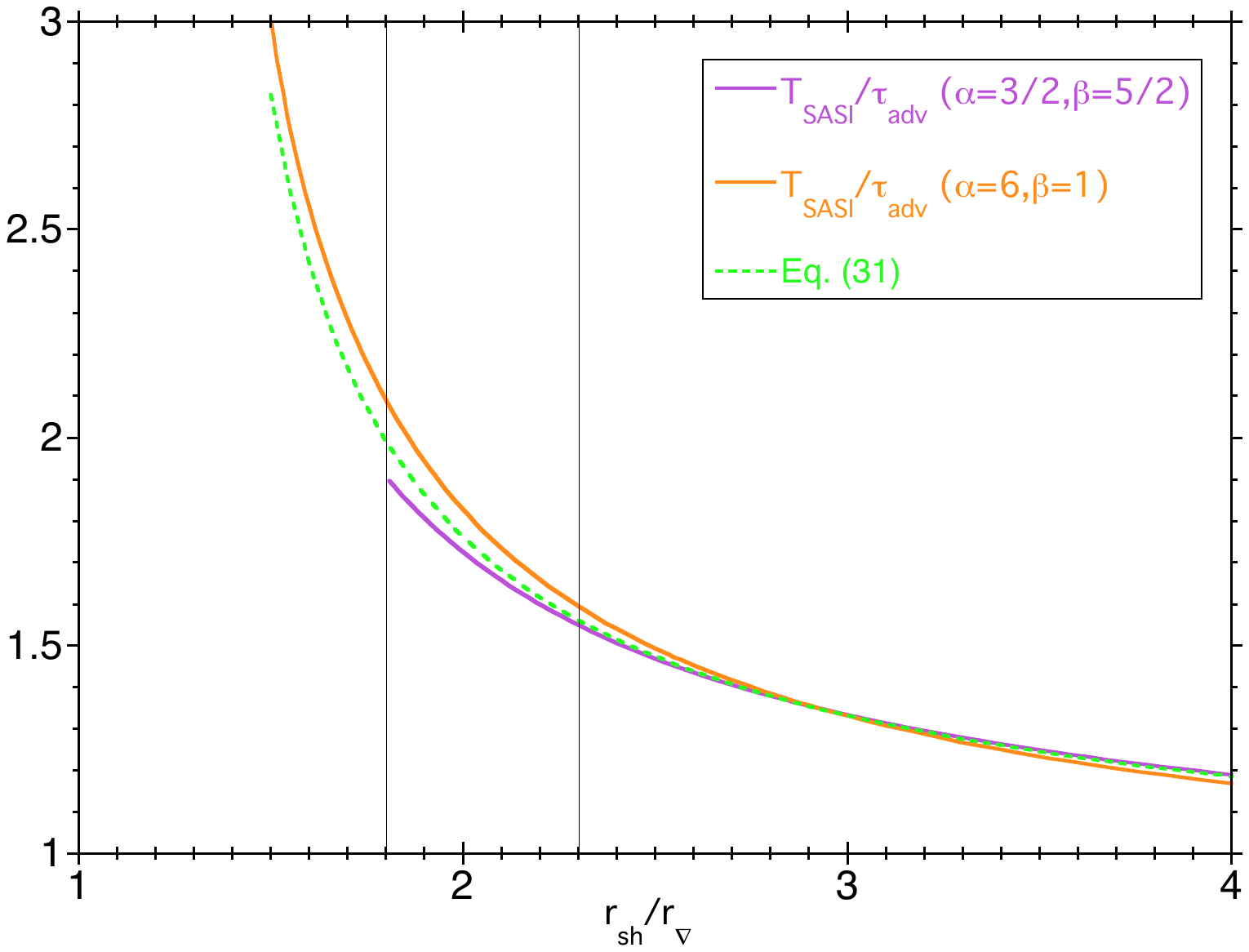}
\caption{Variation of the ratio $T_{\rm SASI}/\tau_{\rm adv}^{\rm adiab}$ associated to the fundamental SASI mode $\ell=1$, calculated for the cooling parameters $(3/2,5/2)$ (purple) and $(6,1)$ (orange) and displayed as functions of $r_{\rm sh}/r_\nabla$. The vertical lines highlight the range of $r_{\rm sh}/r_{\rm ns}$ in the analysis of model s25 in \cite{Muller_Janka2014}. The analytic fitting formula (\ref{eqfitwr}) is displayed with a green dashed line.}
\label{fig_fitphiQ}
\end{figure}
We remark on Fig.~\ref{fig_fitphiQ} that the oscillation period $T_{\rm SASI}\equiv 2\pi/\omega_r$ of the fundamental $\ell=1$ SASI mode can be significantly longer than the approximate advection time $\tau_{\rm adv}^{\rm adiab}$ for a small ratio $r_{\rm sh}/r_\nabla$. The relationship between these two timescales is further discussed in a simplified model in Sect.~\ref{Sect_freq_SASI_MJ}. \\

The modest impact of the specificities of the parametrized cooling process on the fundamental SASI eigenfrequency encourages us to apply our simple model to a more realistic physical model of core-collapse in Sect.~\ref{MJ_SASI}. This also invites us to look for a deeper analytic understanding of the SASI mechanism using an adiabatic approximation, in Sect.~\ref{sect_adiabatic}.

\subsection{Comparison of the perturbative model with the empirical formula proposed by \cite{Muller_Janka2014}}
\label{MJ_SASI}
\begin{figure}
\centering
\includegraphics[width=\columnwidth]{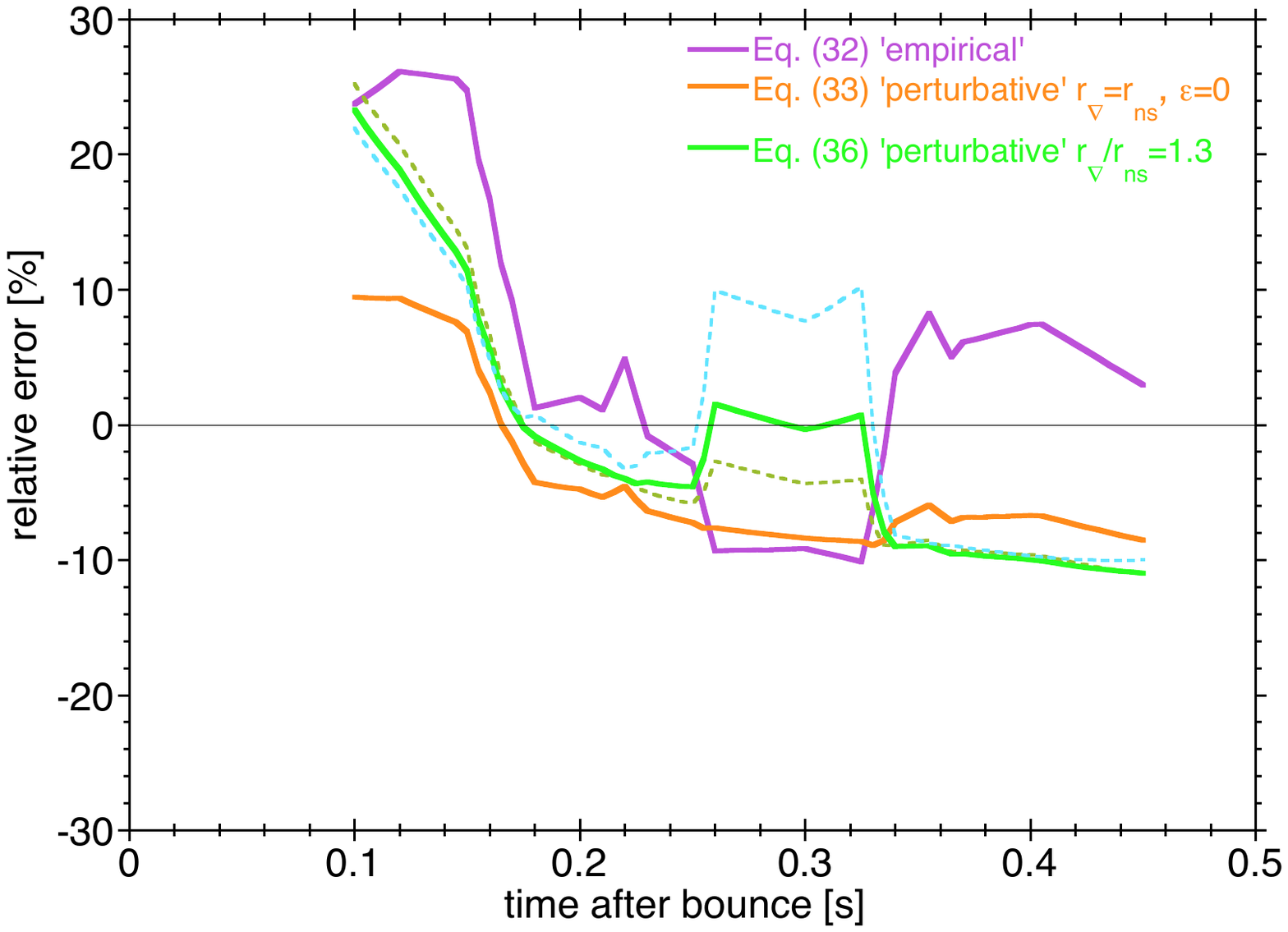}
\caption{
Estimate of the relative error between the oscillation period of the neutrino signal in model s25 in \cite{Muller_Janka2014} 
and in analytical estimates. The empirical formula (\ref{TSASI_MJ2014}) is shown with a purple line. The result of the perturbative calculation is shown for $r_\nabla=r_{\rm ns}$ without dissociation (orange line, Eq.~\ref{improved_TSASI}), and for $r_\nabla=1.3r_{\rm ns}$ with dissociation prescribed by Eq.~(\ref{disso0.5}) (green line, Eq.~\ref{improved_TSASI_disso}). The sensitivity to the estimate of $r_\nabla$ is shown with dashed lines for $r_\nabla=1.25r_{\rm ns}$ (khaki) and $r_\nabla=1.35r_{\rm ns}$ (blue)
}
\label{fig_TSASI_improved}
\end{figure}
\begin{figure}
\centering
\includegraphics[width=\columnwidth]{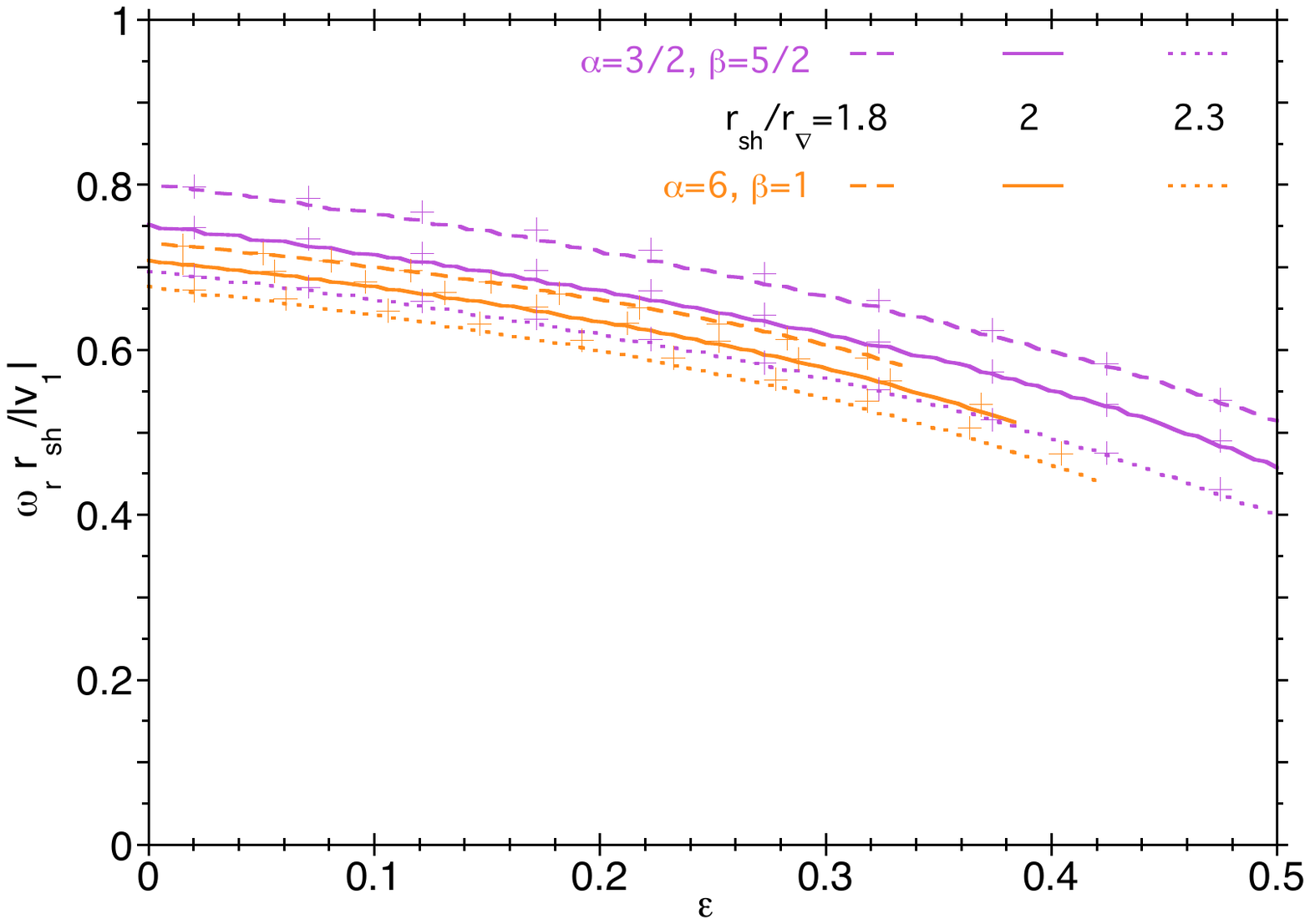}
\caption{Effect of dissociation, measured by $\varepsilon$, on the oscillation frequency of SASI for $r_{\rm sh}/r_\nabla=1.8$ (dashed lines), $1$ (solid lines) and $2.3$ (dotted lines) for $(\alpha,\beta)=(3/2,5/2)$ (purple lines) and $(\alpha,\beta)=(6,1)$ (orange lines). The analytical formula (\ref{omegr_disso}) is indicated with crosses. }
\label{fig_wr_disso}
\end{figure}
The adiabatic approximation of the advection time was also used in the empirical formula (33) describing the oscillation period $T_{\rm SASI}$ in the core-collapse simulation of a $25M_{\odot}$ progenitor \citep{Muller_Janka2014}, inspired by the physics of the advective-acoustic cycle. $r_\nabla$ was approximated with $r_{\rm ns}$ and the time variation of the mass of the proto-neutron star was neglected: 
\begin{eqnarray}
T_{\rm SASI}^{\rm MJ}
&\equiv&
19{\rm ms}\left({r_{\rm sh}\over100{\rm km}}\right)^{3\over2}
\log\left({r_{\rm sh}\over r_{\rm ns}}\right).
\label{TSASI_MJ2014}
\end{eqnarray}
The SASI modulation of the neutrino signal was identified in their model s25 between $t=0.12$s and $t=0.45$s post-bounce. The shock radius $r_{\rm sh}$ decreases from $125$km to $55$km and the ratio $r_{\rm sh}/r_{\rm ns}$ varies between $1.8$ and $2.3$ according to their Figs.~3 and 6. The formula (\ref{TSASI_MJ2014}) captures well the $r_{\rm sh}^{3/2}$-dependence which is the main source of variability of $T_{\rm SASI}$ during the collapse. The overall accuracy of this formula is $\sim 26\%$ for model s25 as shown in Fig.~\ref{fig_TSASI_improved}.\\
The expected variation of $T_{\rm SASI}$ estimated from Eq.~(\ref{eqfitwr}) is as follows:
\begin{eqnarray}
T_{\rm SASI}
&=&
{T_{\rm SASI}^{\rm MJ}\times1.82
\left({1.7M_\odot\over M_{\rm ns}}\right)^{1\over2}
\over 0.56773+0.28628R_1-0.031763R_1^2}.
\label{improved_TSASI}
\end{eqnarray}
According to Fig.~\ref{fig_TSASI_improved} the overall accuracy of formula (\ref{improved_TSASI}) applied to the model s25 is improved to $\sim 10\%$, which is remarkable given the simplicity of the perturbative model which does not involve any adjusted parameter and neglects dissociation at the shock. The mass increase $M_{\rm ns}/M_\odot\sim 1.7-2$ estimated from Fig.~2 in \cite{Muller_Janka2014} contributes to a $8\%$ decrease of $T_{\rm SASI}$. A more physical estimate should also take into account dissociation, which can significantly increase $T_{\rm SASI}$ when $r_{\rm sh}$ is large, and the distinction between $r_{\rm ns}$ and $r_\nabla$, which can significantly decrease $T_{\rm SASI}$ when $r_{\rm sh}/r_{\rm ns}$ is smallest.  \\
Following Fig.~\ref{fig_wr_disso}, the impact of dissociation on the oscillation frequency of SASI is tentatively approximated as 
\begin{eqnarray}
\omega_r(\varepsilon)&=&
\omega_r(0)
-\left({2GM_{\rm ns}\over r_{\rm sh}^3}\right)^{1\over 2}
\varepsilon
\left(0.19817 + 0.74804 \varepsilon\right),
\label{omegr_disso}
\end{eqnarray}
We note that this analytical formula is tested only in the narrow range  $1.8<r_{\rm sh}/r_\nabla<2.3$ and meant as an order of magnitude estimate. The dissociation parameter is approximated according to Fig.~12 in \cite{Huete2018}:
\begin{eqnarray}
\varepsilon&\sim& 0.5\left({r_{\rm sh}\over 150{\rm km}}\right)\left({1.3M_\odot\over M_{\rm ns}}\right)
,\label{disso0.5}
\end{eqnarray}
with a saturation at $\varepsilon\sim 0.5$ due to partial dissociation of $\alpha$-nuclei \citep{Fernandez2009b}. Thus Eq.~(\ref{omegr_disso}) is rewritten as
\begin{eqnarray}
{10{\rm ms}\over T_{\rm SASI}}&=&
{10{\rm ms}\over T_{\rm SASI}^0}
-{1.069\varepsilon
\left(0.19817 + 0.74804 \varepsilon\right)
\over
\left({r_{\rm sh}\over100{\rm km}}\right)^{3\over2}
\left({1.7M_\odot\over M_{\rm ns}}\right)^{1\over2}
}
,
\label{improved_TSASI_disso}
\end{eqnarray}
with
\begin{eqnarray}
{T_{\rm SASI}^0\over 10.42{\rm ms}}&\equiv &
{
\left({r_{\rm sh}\over100{\rm km}}\right)^{3\over2}
\left({1.7M_\odot\over M_{\rm ns}}\right)^{1\over2}
\log\left({r_{\rm sh}\over r_\nabla}\right)
\over 0.56773+0.28628R_1-0.031763R_1^2
}.
\end{eqnarray}
According to Fig.~\ref{fig_TSASI_improved}, taking into account dissociation does improve the accuracy of the empirical formula only for a limited range of prescribed ratio $1.25<r_\nabla/r_{\rm ns}<1.35$. Improving the accuracy of the present perturbative model beyond the $\pm 10\%$ level seems elusive given its many approximations regarding the microphysics, neutrino interactions and gravity. Its main improvement compared to the empirical formula (\ref{TSASI_MJ2014}) is the fact that it contains physically consistent estimates of the impact of the mass, the dissociation at the shock and the SASI phase inferred from the perturbative analysis. Each of them can potentially affect the oscillation period by several $10\%$ according to Eq.~(\ref{improved_TSASI}) and Figs.~\ref{fig_fitphiQ}, \ref{fig_wr_disso}. By neglecting these effects, the empirical formula Eq.~(\ref{TSASI_MJ2014}) is not expected to maintain its $\sim 30\%$ accuracy for other progenitors than s25. The accuracy of Eq.~(\ref{improved_TSASI_disso}) should be tested on other numerical simulations of core collapse, remembering that our prescription for dissociation is very crude.

\section{Adiabatic model} 
\label{sect_adiabatic}

The adiabatic character of the flow refers here to the region between the shock and the inner boundary, while non adiabatic processes associated to nuclear dissociation and neutrino emission are incorporated in the boundary conditions. The production of entropy perturbations by the shock displacement $\Delta\zeta$ is taken into account. Nuclear dissociation across the shock is formally taken into account using the parameter $\varepsilon$ (Eq.~\ref{norm_disso}), but it is set to zero in the illustrative Figures of this Section. The goal of this Section is to gain some analytical understanding on the effect of the physical parameters involved in the SASI mechanism.

\subsection{Explicit expressions in the adiabatic approximation}

With ${\cal L}=0$ the differential equations (\ref{dSdr_main}-\ref{dKdr_main}) describe the advection of perturbations produced by the shock. 
Using the boundary conditions (\ref{dSsh_zeta}-\ref{dKsh}) with $\nabla S=0$:
\begin{eqnarray}
\delta S&=&\delta S_{\rm sh}{\rm e}^{i\omega\int_{\rm sh} {{\rm d}r\over v }},\label{integ_dS}\\
\delta K&=&0.\label{integ_dK}
\end{eqnarray} 
$\delta S_{\rm sh}$ is defined by Eq.~(\ref{dSsh_zeta}) with a simpler expression for $\omega_\Phi$:
\begin{eqnarray}
{\omega_\Phi r_{\rm sh}\over |v_{\rm sh}|}
&\equiv&
{ {1\over 2}{v_1\over\v2}
-g
\over
1-{\v2\over v_{1}}
}.\label{def_omPhi_adiab}
\end{eqnarray}
Using Eq.~(\ref{defdK0}) with $\delta K=0$ and Eq.~(\ref{integ_dS}) gives the explicit expression for $\delta w_\perp$ 
\begin{eqnarray}
\delta w_\perp=-\ell(\ell+1){c^2\over \gamma rv}
\delta S.\label{wperp_explicit}
\end{eqnarray} 
The vorticity associated to the entropy perturbation is also explicitly calculated in Appendix~\ref{App_adiab}:
\begin{eqnarray}
\delta w_r&=&0,\label{dwr_main}\\
\delta w_\theta&=&-{imc^2\over rv\sin\theta}{\delta S\over\gamma},\label{dwtheta_main}\\
\delta w_\varphi&=&{c^2\over rv}{\partial \over\partial\theta}{\delta S\over\gamma}.\label{dwphi_main}
\end{eqnarray} 
Together with Eq.~(\ref{defdK0}), Eqs.~(\ref{dwtheta_main}-\ref{dwphi_main}) demonstrate that $\delta K=0$ can be interpreted as a consequence of the baroclinic production of vorticity.\\
Eq.~(\ref{AFK}) implies that the wave action $\delta f/\omega$ is directly related to $\delta A$:
\begin{eqnarray}
{\delta f\over i\omega}=-{\delta A\over \ell(\ell+1)},
\end{eqnarray}
The transverse velocity components $(\delta v_\theta,\delta v_\varphi)$ are deduced from $\delta A$ using the transverse Euler equations Eqs.~(\ref{ddfdtheta},\ref{ddfdphi}) with Eqs.~(\ref{dwtheta_main},\ref{dwphi_main}). $\delta v_r$ is deduced from Eqs.~(\ref{dvrw_main}) and (\ref{wperp_explicit}):
\begin{eqnarray}
\delta v_r&=&
-{c^2\over \gamma v}\delta S
-{1\over \ell(\ell+1)}{\partial \delta A\over\partial r},\label{dvrdA}
\\
r\delta v_\theta&=&
-{1\over \ell(\ell+1) }{\partial \delta A\over\partial\theta}
,\label{rdvtheta}\\
r\delta v_\varphi&=&-{im\over \ell(\ell+1) }{\delta A\over\sin\theta}
.\label{rdvphi}
\end{eqnarray} 
The adiabatic solution can be viewed as the asymptotic limit of the model with a cooling function when $r_{\rm peak}\sim r_{\rm ns}$, corresponding to $\alpha\ll 1$ and $\beta\gg1$ according to Fig.~\ref{fig_profile_cooling}. For the sake of simplicity we explore the adiabatic solution with an inner boundary defined by $\delta v_r=0$, formulated by Eq.~(\ref{inner_BC}) with $c_{\rm ns}$ defined as the adiabatic sound speed at the inner boundary. This prescription could be improved to better account for non adiabatic processes in the cooling layer.
The adiabatic simulations in \cite{Blondin_Mezzacappa2007} also used $\delta v_r=0$ at the inner boundary. 

\subsection{Quantitative comparison of the eigenfrequencies with and without the region of non-adiabatic cooling}

\begin{figure}
\centering
\includegraphics[width=\columnwidth]{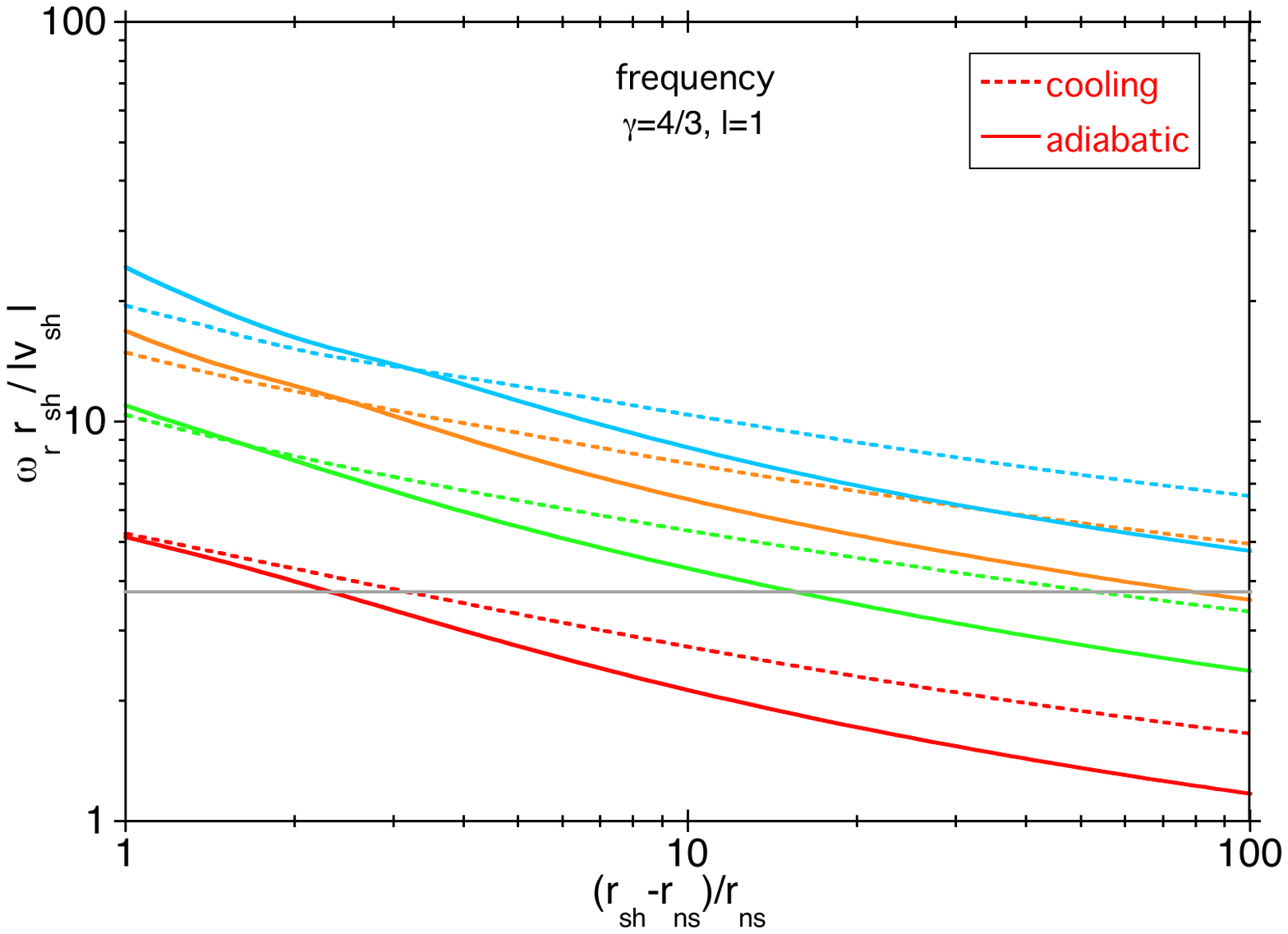}
\includegraphics[width=\columnwidth]{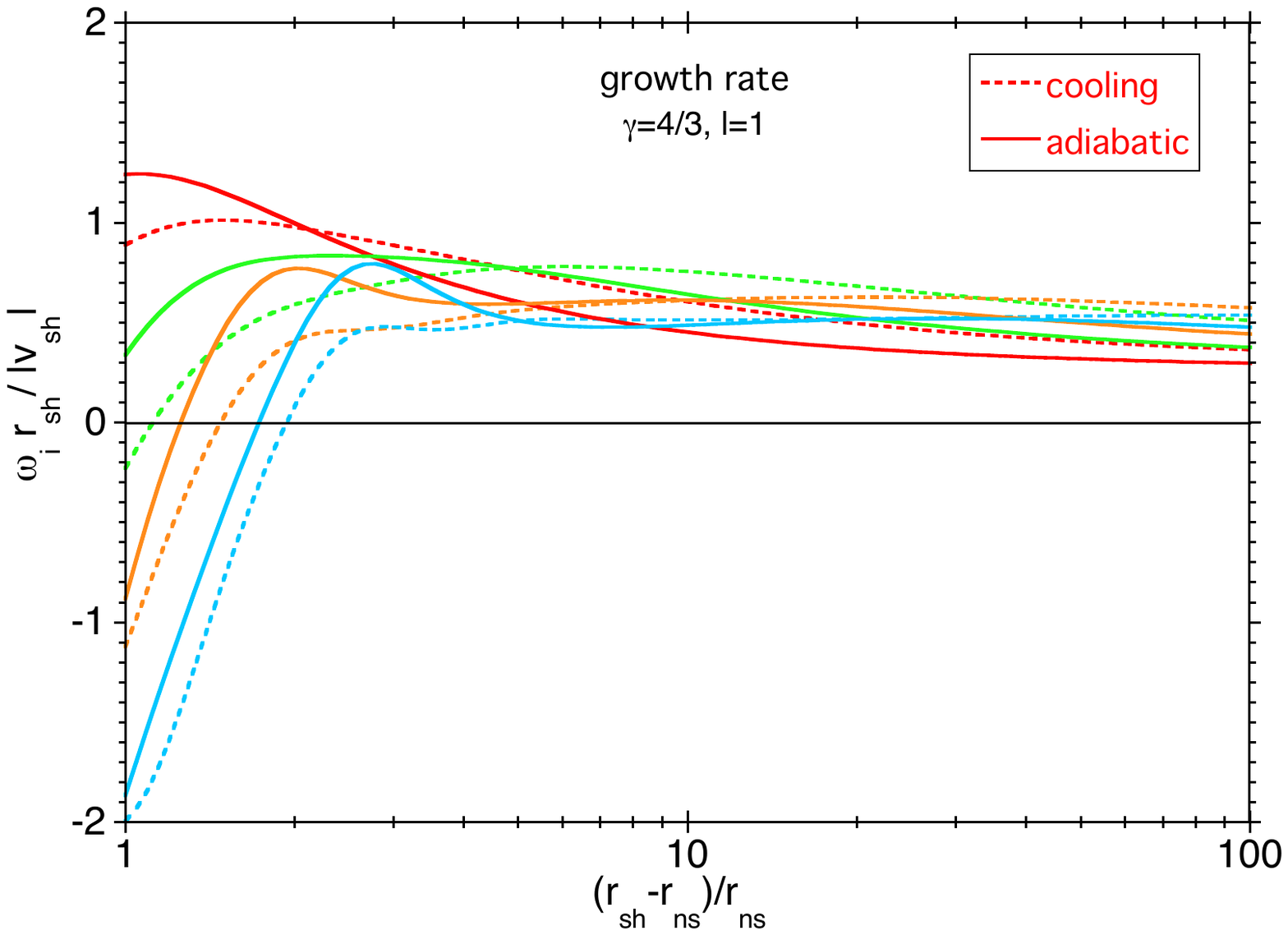}
\caption{Oscillation frequency (upper plot) and growth rate (lower plot) of the modes $\ell=1$ calculated in units of the post-shock frequency $v_{\rm sh}/r_{\rm sh}$, for $\gamma=4/3$, as a function of the shock distance in the model with cooling using $(\alpha,\beta)=(3/2,5/2)$ (dashed lines) and in the adiabatic approximation (solid lines). The fundamental mode (in red) and the first three overtones (green, orange, blue) are displayed. The grey horizontal line in the upper plot indicates the Lamb frequency at the shock. The fundamental mode becomes dominated by higher overtones as its frequency becomes too low for acoustic propagation ($\omega_r<\omega_{\rm Lamb}^{\rm sh}$). }
\label{fig_harmonics}
\end{figure}
The eigenfrequencies corresponding to the adiabatic model are solved numerically and compared in Fig.~\ref{fig_harmonics} to the eigenfrequencies of the non-adiabatic formulation for different values of the shock radius. The overall trends suggest that the main properties of SASI can be qualitatively understood by focusing on adiabatic processes. The fundamental mode is the most unstable one for a small shock radius, and becomes dominated by higher overtones for a larger shock radius. Among the most visible differences with the non-adiabatic model, the adiabatic simplification underestimates the frequency by a factor $\le1.3$ for a large shock radius $r_{\rm sh}\sim 10r_{\rm ns}$. The growth rate of the most unstable mode is overestimated by a factor $\le1.4$ for a small shock radius and underestimated by a factor $\le1.3$ for a very large shock radius. 

\section{Formulation of the SASI mechanism as a self-forced oscillator\label{sect_forced}}

\subsection{Derivation of the second order differential equation}
\label{Sect_derivation_forced}

For the sake of mathematical simplicity we use the radial coordinate $X$ defined as in \cite{Foglizzo2001}, 
\begin{eqnarray}
{\rm d X}&\equiv& {v \over 1-\M^2} {\rm d}r.\label{def_X}
\end{eqnarray} 
The second order differential equation (\ref{eqdiffAcool_main}) is simplified into:
\begin{eqnarray}
\left\lbrack\left({\partial\over\partial X}+{i\omega\over c^2}\right)^2
+
{\omega^2-\omega_{\rm Lamb}^2\over v^2c^2}\right\rbrack
\delta A
=
{\partial\over\partial X}
{r\delta w_\perp\over v}.
\label{forced_dA}
\end{eqnarray} 
Using Eq.~(\ref{wperp_explicit}) the forcing term of Eq.~(\ref{forced_dA}) is thus proportional to the entropy perturbation $\delta S_{\rm sh}$ produced by the shock:
\begin{eqnarray}
\left\lbrack\left({\partial\over\partial X}+{i\omega\over c^2}\right)^2
+
{\omega^2-\omega_{\rm Lamb}^2\over v^2c^2}\right\rbrack
{\delta A\over \ell(\ell+1)}
=
-{\cal F}_{S}\delta S_{\rm sh}
,\label{forced_oscillator}\\
{\cal F}_{S}\equiv {\partial \over\partial X}
{{\rm e}^{i\omega\int_{\rm sh} {{\rm d}r\over v}}\over\gamma\M^2}
.\label{def_FS}
\end{eqnarray} 
We show in Appendix~\ref{App_adiab} that the forced oscillator equation 
can be rewritten as follows, using boldface for vector quantities:
\begin{eqnarray}
\left\lbrack\left({\partial\over\partial X}+{i\omega\over c^2}\right)^2
+
{\omega^2-\omega_{\rm Lamb}^2\over v^2c^2}\right\rbrack
{\bf \delta L}
&=&{\partial\over\partial X}
{r{\bf \delta w}\over v}
,\label{diffspheriquevect}\\
&=&{\partial\over\partial X}
{rv{\bf \delta w}\over c^2\M^2},
\end{eqnarray} 
where ${\bf \delta L}\equiv {\bf r}\times {\bf \delta v}$ is the perturbed specific angular momentum vector and ${\bf \delta w}\equiv\nabla\times {\bf \delta v}$ is the perturbed vorticity vector. We note from Eqs.~(\ref{dwtheta_main}-\ref{dwphi_main}) that $rv|\delta w|/c^2$ is conserved when advected.
The relation between $\delta S_{\rm sh}$ and $\delta A_{\rm sh}$ is deduced from Eqs.~(\ref{dAsh}) and (\ref{dSsh_zeta}):
\begin{eqnarray}
\ell(\ell+1){c_{\rm sh}^2\delta S_{\rm sh}\over \gamma\delta A_{\rm sh}}&=&
-\left(1-{\v2\over v_{1}}\right)
 (i\omega+\omega_\Phi)
.
\end{eqnarray} 
The components of transverse vorticity at the shock are related to the components of perturbed specific angular momentum according to Eqs.~(\ref{dwtheta_main}-\ref{dwphi_main}) and Eqs.~(\ref{rdvtheta}-\ref{rdvphi}):
\begin{eqnarray}
\left({rv\delta w_\theta\over  \delta L_\theta}\right)_{\rm sh}=
\left({rv\delta w_\varphi\over  \delta L_\varphi}\right)_{\rm sh}=
-\ell(\ell+1){c_{\rm sh}^2\delta S_{\rm sh}\over \gamma\delta A_{\rm sh}}.
\end{eqnarray} 
The boundary conditions at the shock and at the inner boundary are written in Appendix~\ref{App_adiab} as follows:
\begin{eqnarray}
{\partial\delta A\over\partial X}_{\rm sh}={i\omega  \over v_{\rm sh}^2}\delta A_{\rm sh}\left\lbrack
1-2{\v2\over v_1}
+(\gamma-1)\M_{\rm sh}^2
\left(1-{\v2\over v_{1}}\right)
\right\rbrack
\nonumber\\
-\left\lbrack1+(\gamma-1)\M_{\rm sh}^2\right\rbrack
{\delta A_{\rm sh}\over r_{\rm sh}v_{\rm sh}}
\left({v_1\over 2\v2}
-2\right),
\label{dYdXsh_0}
\\
{\partial\delta A\over\partial X}_{\rm ns}={i\omega\over   c_{\rm ns}^2}\delta A_{\rm ns}
-\ell(\ell+1)
 {1-\M^2_{\rm ns}\over\M^2_{\rm ns}}{\delta S_{\rm sh}\over\gamma}{\rm e}^{i\omega\int_{\rm sh}^{\rm ns}{{\rm d}X\over v^2}}
.\label{lowerbc_0}
\end{eqnarray} 
They can also be written with the variables $r\delta v_\theta,r\delta v_\varphi$ and $\delta w_\theta,\delta w_\varphi$, using  Eqs.~(\ref{dwtheta_main}-\ref{dwphi_main}) and~(\ref{rdvtheta}-\ref{rdvphi}). Equations~(\ref{forced_oscillator}) and (\ref{diffspheriquevect}) are thus equivalent for $\ell\ge1$. The same advective-acoustic cycle can either be considered as an entropic-acoustic cycle or as a vortical-acoustic cycle. \\
We define a new perturbative variable $\delta Y$ as follows:
\begin{eqnarray}
\delta Y&\equiv&
{\delta f\over i\omega}
{\rm e}^{i\omega\int_{\rm sh} 
{{\rm d}X\over c^2}}
=-{\delta A\over\ell(\ell+1)}
{\rm e}^{i\omega\int_{\rm sh} 
{{\rm d}X\over c^2}}
.\label{def_Y}
\end{eqnarray} 
$Y_0$ is defined as the solution of the homogeneous equation associated to Eq.~(\ref{forced_oscillator}) with $\delta S=0$ and $\delta K=0$, and satisfying the inner boundary condition (\ref{lowerbc_0}):
\begin{eqnarray}
\left\lbrace{\partial^2\over\partial X^2}+
{\omega^2-\omega_{\rm Lamb}^2\over v^2c^2}
\right\rbrace Y_0
=0,\label{homogeneous_Y0}\\
{\partial Y_0\over\partial X}_{\rm ns}={i\omega \over   c_{\rm ns}^2} Y_0(r_{\rm ns}).\label{homogeneous_LBC}
\end{eqnarray} 
We refer to $Y_0$ as the acoustic structure of the post-shock cavity modified by the radial velocity, in the absence of interaction with advected perturbations. The structure of the acoustic equation and its modification by the advection velocity $v$ is more easily recognized when rewriting Eq.~(\ref{homogeneous_Y0}) with the variable $r$:
\begin{eqnarray}
\left\lbrace
{\partial^2\over\partial r^2}+
\left({\partial\log\over\partial r}{1-\M^2\over v}\right)
{\partial\over\partial r}+
{\omega^2-\omega_{\rm Lamb}^2\over c^2(1-\M^2)^2}
\right\rbrace Y_0
=0.\label{homogeneous_Y0r}
\end{eqnarray} 
Identifying the physics of SASI with a forced oscillator enables us to evaluate the efficiency of the coupling depending on two effects:
\par(i) the amplitude of the forcing term ($\propto 1/\M^2$ in Eq.~\ref{def_FS}),
\par(ii) the phase match between the forcing term (advected wavelength) and the oscillator (acoustic wavelength).\\
The forcing amplitude is strongest where $\M$ is smallest, in the close vicinity of the proto-neutron star, but a strong phase mixing is expected there due to decrease of the radial wavelength ($\propto 2\pi|v|/\omega_r$) of advected perturbations. A trade-off between the effects (i) and (ii) favours advected perturbations with a low enough frequency, coupled to the acoustic structure in a region far enough from the proto-neutron star.\\
This description as a forced oscillator had been proposed in the context of radial Bondi accretion accelerated into a black hole \citep{Foglizzo2001, Foglizzo2002} and in a planar adiabatic toy model of SASI with a localized region of feedback (F09). The present work is the first formulation where the radially extended character of the advective-acoustic coupling is taken into account in a shocked decelerated accretion flow in spherical geometry. \\
The quantitative evaluation of the coupling efficiency is analyzed in the next Sect.~\ref{sec_integral_eq} using a classical resolution of the forced oscillator with the Wronskian method. 

\subsection{Integral equation defining the eigenfrequencies\label{sec_integral_eq}}

We derive in Appendix~\ref{integral_dispersion} the integral equation defining the eigenfrequencies, which is equivalent to the full differential system and boundary conditions. It is formulated here with the variable $r$:
\begin{eqnarray}
a'_1 Y_0^{\rm sh}
+a'_2r_{\rm sh}  \left({\partial Y_0\over\partial r}\right)_{\rm sh}
=-\M_{\rm sh}^2{\rm e}^{\int_{\rm sh}^{\rm ns}{i\omega\over v}{{\rm d}r\over 1-\M^2}} Y_0^{\rm ns}
\nonumber\\
-\int_{\rm ns}^{\rm sh} 
{\partial \over\partial r}\left(
 Y_0 
{\rm e}^{\int_{\rm sh} {i\omega\M^2\over 1-\M^2}{{\rm d}r\over v}}
\right)
{\M_{\rm sh}^2\over\M^2}
{\rm e}^{\int_{\rm sh} {i\omega\over v}{\rm d}r}
{\rm d}r,\label{single_disp}
\end{eqnarray}
with $a'_1,a'_2$ defined by:
\begin{eqnarray}
a'_1
&\equiv&
(\gamma-1)\M_{\rm sh}^2+
{i\omega\over i\omega+\omega_\Phi}
{
{\v2\over v_1}
\over
1-{\v2\over v_{1}}
}\label{def_a1p}
,\\
a'_2&\equiv&
-
{
1-\M_{\rm sh}^2
\over
\left(1-{\v2\over v_{1}}\right)
{(i\omega+\omega_\Phi) r_{\rm sh}\over \v2}
}
.\label{def_a2p}
\end{eqnarray}
The first two terms associated to $Y_0$ on the left-hand-side of Eq.~(\ref{single_disp}) are acoustic, marginally modified by advection. The terms on the right-hand-side involve the phase oscillations of the advected perturbations. Despite the integration by part we recognize in the integral the forcing term of Eq.~(\ref{def_FS}) multiplied by the derivative of the homogeneous solution $Y_0$.
As expected in Sect.~\ref{Sect_derivation_forced} for the classical problem of a forced harmonic oscillator, this integral characterizes the efficiency of the forcing which depends on both the amplitude profile of the forcing $\delta {\cal F}$ and the matching of its phase compared to the phase of the oscillator $Y_0$. An analytic approximation of this integral equation is obtained in Sect.~\ref{sect_approx_eigen} in the asymptotic regime where the acoustic radial structure is non-oscillatory.\\
We checked numerically that the eigenfrequencies satisfying the single equation~(\ref{single_disp}) are strictly the same as obtained in Fig.~\ref{fig_harmonics} from the solution of the fourth order adiabatic differential system (\ref{dAdr_main}-\ref{dKdr_main}) with ${\cal L}=0$, with the boundary conditions defined by Eq.~(\ref{inner_BC}) and 
Eqs.~(\ref{dAsh}-\ref{dKsh}) with $\nabla S=0$.

\section{Analytic estimate for a large shock radius}
\label{sect_approx_eigen}
\subsection{Analytic approximations\label{Sect_approx}}

\subsubsection{Stationary flow}

We focus on the case of a strong shock $\M_1\gg1$ and large shock radius $r_{\rm sh}\gg r_{\rm ns}$, with the post-shock energy density described by the Bernoulli equation (\ref{eq_euler_stat}) is dominated by the enthalpy and the gravitational contributions:
\begin{eqnarray}
{c^2\over\gamma-1}
&\sim& {GM_{\rm ns}\over r}.\label{Bersimpl_m}
\end{eqnarray}  
The kinetic energy density associated to the radial velocity is a minor contribution which is even more negligible when the photodissociation across the shock is taken into account. The parameter $\varepsilon$ impacts the post-shock Mach number and velocity according to Eqs.~(\ref{varepsilon}-\ref{v2v1}):
\begin{eqnarray}
\M_{\rm sh}^2&\sim&{\gamma-1\over2\gamma}(1-\varepsilon),\label{Machsimpl_m}\\
{\v2\over v_1}&\sim&{(\gamma-1)(1-\varepsilon) \over 2+(\gamma -1)(1-\varepsilon)}
\le{\gamma-1\over \gamma+1}\label{v1v2simpl_m},\\
{c_{\rm sh}^2r_{\rm sh}\over GM_{\rm ns}}
&\sim& 
{4\gamma(\gamma-1)(1-\varepsilon) \over \left\lbrack 2+(\gamma -1)(1-\varepsilon)\right\rbrack^2}.
\label{c2G_M_m}
\end{eqnarray}  
Eq.~(\ref{Bersimpl_m}) allows for a power law approximation of the radial velocity profile with an exponent $\alpha_v$ deduced from Eq.~(\ref{mass_cons}). It depends on both the adiabatic index and the geometry $g=2$:
\begin{eqnarray}
\alpha_v&\equiv&{1\over\gamma-1}-g,\label{def_alphav}\\
v&=&v_{\rm sh}\left({r\over r_{\rm sh}}\right)^{\alpha_v}\label{approx_v},\\
{\M_{\rm sh}\over \M}&\sim&\left({r_{\rm sh}\over r}\right)^{\alpha_v+{1\over 2}}.\label{approx_M}
\end{eqnarray} 
We note that Eq.~(\ref{Bersimpl_m}) is valid only for $r_{\rm ns}\le r\ll r_{\rm sh}$, and becomes inaccurate in the vicinity of the shock where the sound speed is particularly sensitive to photodissociation as seen in Eq.~(\ref{c2G_M_m}). 

\subsubsection{Acoustic structure of the oscillator}

\begin{figure}
\centering
\includegraphics[width=\columnwidth]{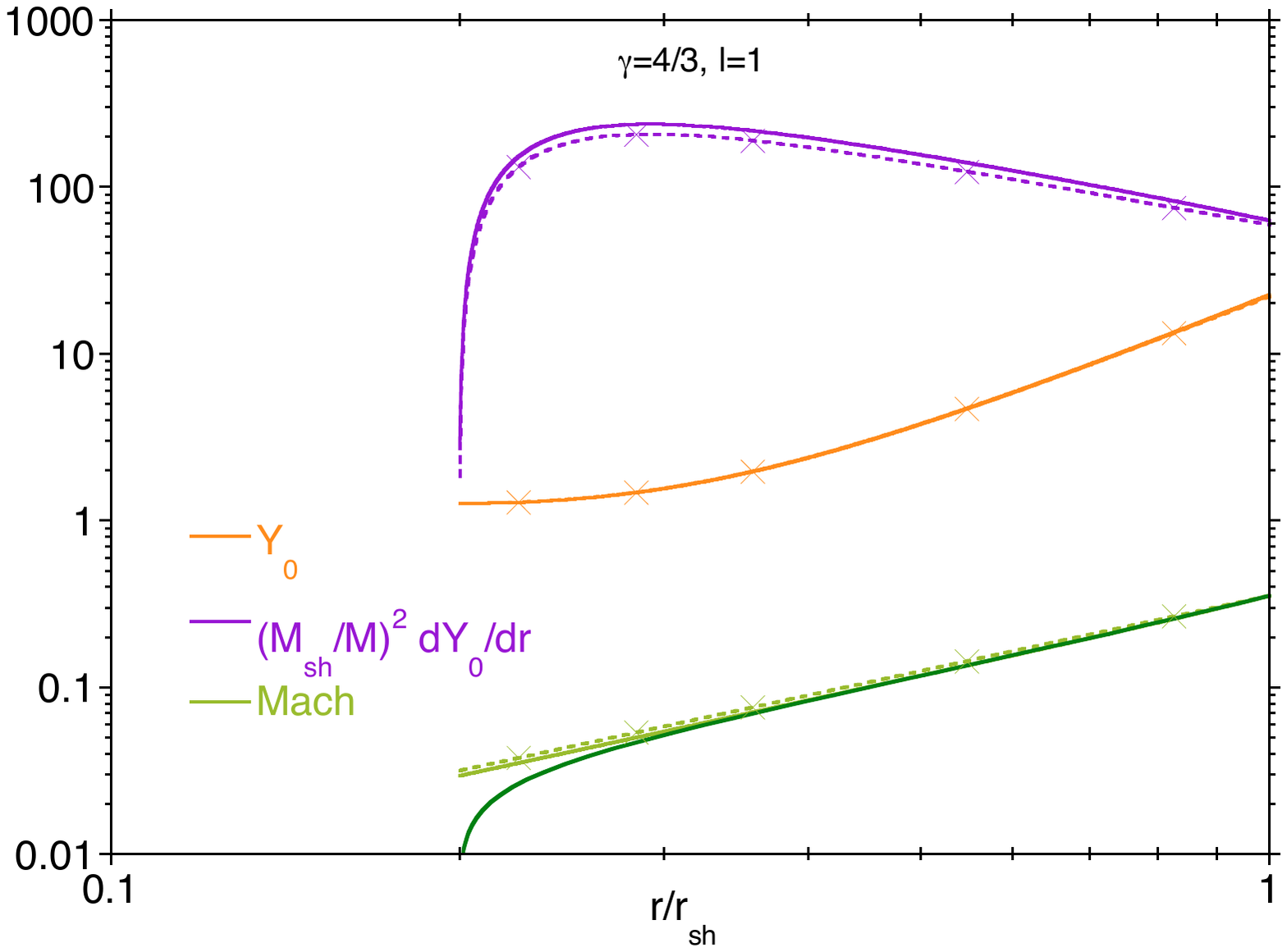}
\caption{Solution of the homogeneous equation $Y_0$ associated to the fundamental mode $\ell=1$ (solid orange line) for $r_{\rm sh}/r_{\rm ns}=5$. It is very well approximated analytically by Eq.~(\ref{approx_Y0}) (dashed orange line with crosses). 
The radial profile of the amplitude of the forcing term in Eq.~(\ref{single_disp}) (purple solid line) is compared to its analytical approximation (dashed purple line with crosses) using Eqs.~(\ref{approx_M}) and (\ref{dY0dr}). 
The Mach number profile in the adiabatic model (green solid line) is compared to its analytical approximation (Eqs.~\ref{approx_M}, green dashed line with crosses). The Mach profile of the flow including cooling is shown for reference (dark green line). 
}
\label{fig_statY0}
\end{figure}
Taking advantage of the low frequency of the fundamental mode for $\ell\ge1$ we approximate in Appendix~\ref{app_approx} the homogeneous solution $Y_0$ as a linear combination of power laws independent of the frequency for $\omega\ll\omega_{\rm Lamb}$
\begin{eqnarray}
Y_0&\sim&
\left({r\over r_{\rm ns}}\right)^{\a}\left\lbrack
\left({r_{\rm ns}\over r}\right)^{\b}
+
{\b-\a
\over
\b+\a}
\left({r\over r_{\rm ns}}\right)^{\b}
\right\rbrack,
\label{approx_Y0}
\\
{\partial Y_0\over\partial r}&\sim&
{\b-\a\over r_{\rm ns}}
\left({r\over r_{\rm ns}}\right)^{\a-1}
\left\lbrack
\left({r\over r_{\rm ns}}\right)^{\b}
-
\left({r_{\rm ns}\over r}\right)^{\b}
\right\rbrack.\label{dY0dr}
\end{eqnarray}
where the exponents $\a,\b$ are defined by:
\begin{eqnarray}
\a&\equiv&{\alpha_v+1\over 2},\label{def_alpha1}\\
\b&\equiv&\left\lbrack
\left({{\alpha_v}+1\over 2}\right)^2+\ell(\ell+1)
\right\rbrack^{1\over 2}.\label{def_alpha2}
\end{eqnarray}
The exponents $\pm \b$ correspond to the two components of the acoustic structure $Y_0$, with the component decreasing inward ($+\b$) associated to the evanescent profile of pressure perturbations at the shock, and the component decreasing outward ($-\b$) associated to the evanescent profile of pressure perturbations at the inner boundary.\\
This approximate solution of $Y_0$ satisfies the lower boundary condition only in the asymptotic limit $r_{\rm ns}\ll r_{\rm sh}$.

\subsubsection{Forcing by advected perturbations}

We use the following approximation of the advection time $\tau_{\rm adv}(r)$ profile:
\begin{eqnarray}
\tau_{\rm adv}(r)&\equiv& \int_{\rm sh}^r{{\rm d}r\over v}\label{tadv_exact},\\
&\sim&
{1\over \alpha_v-1}{r_{\rm sh}\over |v_{\rm sh}|}\left\lbrack \left({r_{\rm sh}\over r}\right)^{\alpha_v-1}-1\right\rbrack\;\;{\rm if}\;\;\alpha_v\ne1,\label{approx_I1}\\
&\sim&
{r_{\rm sh}\over |v_{\rm sh}|}
\log{r_{\rm sh}\over r}\;\;{\rm if}\;\;\alpha_v=1,\label{tadvsph}
\end{eqnarray} 
In particular Eq.~(\ref{tadvsph}) is applicable for $\gamma=4/3$. In this regime the advection time $\tau_{\rm adv}$ is asymptotically dominated by the inner region. A numerical calculation indicates that this analytical formula slightly underestimate the adiabatic advection time by less than $6\%$ in the range $1<r_{\rm sh}/r_{\rm ns}<100$.\\
We check the accuracy of our approximation of ${\cal M}$ and $Y_0$ in Fig.~\ref{fig_statY0}.

\subsection{Analytic expression of the fundamental eigenfrequency}
\label{Sect_analytic_eigen}

We approximate Eq.~(\ref{single_disp}) using Eq.~(\ref{approx_M}) and Eq.~(\ref{approx_Y0}) with $\M^2\ll1$ except near the shock.
We use the approximations (\ref{approx_Y0}) and (\ref{dY0dr}) of $Y_0$ and $\partial Y_0/\partial r$ and note that $(\partial \log Y_0/\partial \log r)_{\rm sh}\sim 1/(\b+\a)$ with $\a,\b$ defined by Eqs.~(\ref{def_alpha1}) and (\ref{def_alpha2}). We define $N\equiv a_2'+a_1'/(\b+\a)$  and introduce the normalized eigenfrequency $Z\equiv i\omega r_{\rm sh}/|v_{\rm sh}|$:
\begin{eqnarray}
\N\left(Z\right)\equiv 
{\gamma-1\over\b+\a}\M_{\rm sh}^2+
{
1-\M_{\rm sh}^2-{Z\over\b+\a}
{\v2\over v_1}
\over
\left(1-{\v2\over v_{1}}\right)
\left(Z+{\omega_\Phi r_{\rm sh}\over |v_{\rm sh}|}\right)
}.\label{def_komeg}
\end{eqnarray}
Using the variable $x\equiv r/r_{\rm ns}$, for $\ell\ge1$ the eigenfrequency equation (\ref{single_disp}) is approximated by 
\begin{eqnarray}
-x_{\rm sh}^{3\a-\b-1}\int_{1}^{x_{\rm sh}}
\left(
x^{\b}
-
x^{-\b}
\right)
{\rm e}^{i\omega\int_{\rm sh} {{\rm d}r\over v}}
{{\rm d}x\over x^{3\a}}
=\nonumber\\
\N\left({i\omega r_{\rm sh}\over |v_{\rm sh}|}\right)
+{2\b
\over
\ell(\ell+1)}
{\M_{\rm sh}^2\over x_{\rm sh}^{\a+\b}}
{\rm e}^{i\omega \tau_{\rm adv}^{\rm ns} }
.
\label{simplest_dispnorot_approx}
\end{eqnarray}
\begin{figure}
\centering
\includegraphics[width=\columnwidth]{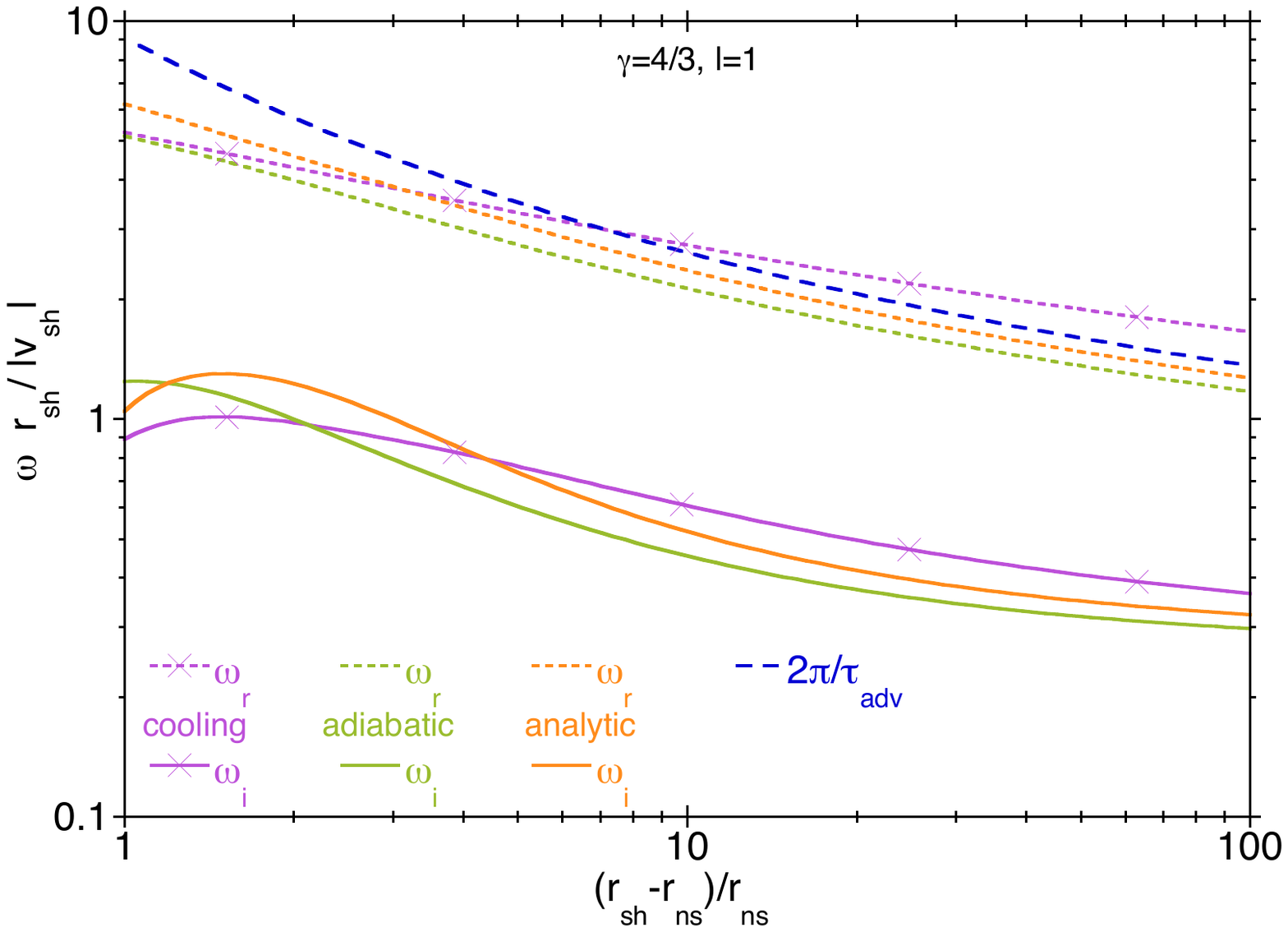}
\caption{Eigenfrequency of the adiabatic solution (green lines) based on the integral formulation (\ref{single_disp}) compared to its integrated approximation (\ref{implicit_analytic}) (orange lines) based on the approximate description of the flow profile $v,\M$ and acoustic structure $Y_0$. The estimate of the oscillation frequency $2\pi/\tau_{\rm adv}^{\rm ns}$ neglecting the phase of ${\cal Q}$ is shown as a dashed blue line. For reference the eigenfrequency of the flow with cooling using $(\alpha,\beta)=(3/2,5/2)$ is also shown (purple lines with crosses).}
\label{fig_sphernorot}
\end{figure}

\begin{figure}
\centering
\includegraphics[width=\columnwidth]{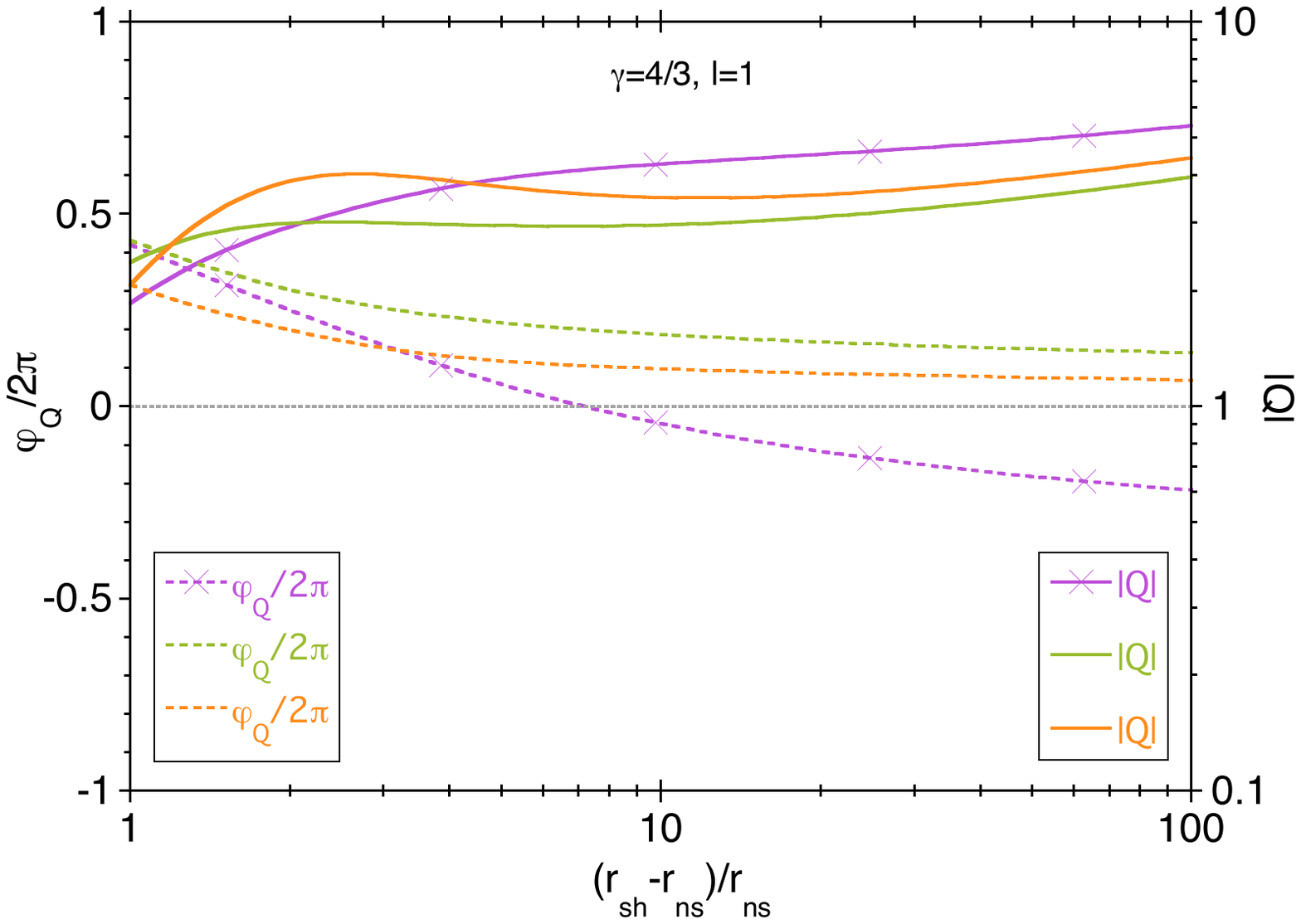}
\caption{Amplitude $|{\cal Q}|$ (solid lines) and normalized phase $\varphi_{\cal Q}/2\pi$ (dashed lines) of the complex amplification factor associated through Eqs.~(\ref{def_absQ}-\ref{def_phQ}) to the fundamental eigenfrequency $\omega$ calculated  with cooling (purple lines with crosses) and in the adiabatic approximation (green lines), using the same adiabatic estimate of the advection time $\tau_{\rm adv}^{\rm ns}$ from Eq.~(\ref{tadvsph}). 
The orange line shows $|{\cal Q}|$ and $\varphi_{\cal Q}/2\pi$
deduced from the analytic estimate (orange lines) using Eq.~(\ref{implicit_analytic2}) with $\omega$ satisfying Eq.~(\ref{implicit_analytic}).}
\label{fig_phase}
\end{figure}

The approximation of the eigenfrequency with $\gamma=4/3$ allows for an explicit calculation of the integral involved in Eq.~(\ref{simplest_dispnorot_approx}). 
The following equation defining the eigenfrequencies for $\ell\ge1$ is expressed using the complex amplification factor ${\cal Q}(\omega)$ per 
advective-acoustic cycle:
\begin{eqnarray}
{\cal Q}(Z)\equiv
{2\b \left({r_{\rm sh}\over r_{\rm ns}}\right)^{2-\b}
\left\lbrace
1+\left\lbrack (Z+2)^2-\b^2 \right\rbrack {\M_{\rm sh}^2\over \ell(\ell+1) x_{\rm sh}^3}
\right\rbrace
\over
\left\lbrack
1
-\left(Z+2-\b\right)
\N
\right\rbrack
\left(Z+2+\b\right)
-{Z+2-\b\over x_{\rm sh}^{2\b}}
}
,\label{implicit_analytic2}
\\
{\cal Q}\left({i\omega r_{\rm sh}\over |v_{\rm sh}|}\right)
{\rm e}^{i\omega \tau_{\rm adv}^{\rm ns} }=1,\label{implicit_analytic}
\end{eqnarray}
where $\b^2=1+\ell(\ell+1)$ and the advection time $\tau_{\rm adv}^{\rm ns}$ from the shock to the inner boundary is approximated by Eq.~(\ref{tadvsph}).
The numerical solution of this equation is compared to the exact solution of Eq.~(\ref{single_disp}) in Fig.~\ref{fig_sphernorot}. 

The good agreement obtained with the analytic expression (\ref{implicit_analytic}) demonstrates that the approximation of the flow profile $v$, $\M$ and the approximation of the homogeneous solution $Y_0$ is sufficient to capture the essence of the instability in the adiabatic model and identify the leading contributions in the integral expression (\ref{single_disp}).
The frequency is overestimated by up to $20\%$ and the growth rate is overestimated by up to $26\%$. The region of radial propagation of acoustic waves was neglected for the sake of simplicity in the analytical approximation (\ref{approx_Y0}) of $Y_0$, without much consequences on the approximation of the fundamental frequency $\omega_r$ because the radial extension of this region is modest compared to the region affecting the phase of advected perturbations.\\
The larger inaccuracy $\sim 76\%$ of the simple formula $\omega_r\sim 2\pi/ \tau_{\rm adv}^{\rm ns}$, also shown in Fig.~\ref{fig_sphernorot}, is mainly due to neglecting the frequency dependence of the phase of ${\cal Q}$ noted $\varphi_{\cal Q}$. The complex Eq.~(\ref{implicit_analytic}) is equivalent to the following set of two real equations:
\begin{eqnarray}
\omega_i&=&{{\rm log}|{\cal Q}| \over \tau_{\rm adv}^{\rm ns}},\label{wi_Q}\\
\omega_r&=&{2\pi-\varphi_{\cal Q}\over \tau_{\rm adv}^{\rm ns}}.\label{wr_Q}
\end{eqnarray}

\subsection{Oscillation period of the advective-acoustic cycle in the adiabatic approximation}
\label{Sect_freq_SASI_MJ}

Comparing $\omega_r$ and $2\pi/\tau_{\rm adv}^{\rm ns}$ in Fig.~\ref{fig_sphernorot}, it is particularly striking that the oscillation period $T_{\rm SASI}\equiv 2\pi/\omega_r$ is significantly longer than the adiabatic advection time $\tau_{\rm adv}^{\rm ns}$.
This contrasts with the simulations in \cite{Scheck2008} where $T_{\rm SASI}\sim \tau_{\rm adv}^{\rm ns}$, in line with the analytic toy model of F09. Not only does $T_{\rm SASI}$ differ from $\tau_{\rm adv}^{\rm ns}$ by more than $76\%$, but $T_{\rm SASI}$ is actually $76\%$ {\it longer} than the longest advection timescale $\tau_{\rm adv}^{\rm ns}$, excluding the possibility that the oscillation period could coincide with any advection time.
As explained in \cite{Scheck2008}, the relation between $T_{\rm SASI}$ and $\tau_{\rm adv}^{\rm ns}$ does depend on the phase $\varphi_{\cal Q}$ of the complex efficiency ${\cal Q}$. \\
The adiabatic approximation teaches us that even in a simple adiabatic model the phase shift $\varphi_{\cal Q}$ should not be neglected a priori. This was also true in the calculation of Sect.~\ref{MJ_SASI} with non-adiabatic cooling.
The value of $|{\cal Q}|$ and $\varphi_{\cal Q}$ can be deduced from $\omega$ and $\tau_{\rm adv}^{\rm ns}$ from Eqs.~(\ref{wi_Q}-\ref{wr_Q}):
\begin{eqnarray}
|{\cal Q}| &=& {\rm e}^{ \omega_i \tau_{\rm adv}^{\rm ns}},\label{def_absQ}\\
\varphi_{\cal Q}&=& 2\pi-\omega_r\tau_{\rm adv}^{\rm ns}.\label{def_phQ}
\end{eqnarray}
This characterisation of the fundamental eigenfrequency is calculated in Fig.~\ref{fig_phase} in the adiabatic approximation and compared to the analytic estimate deduced from Eqs.~(\ref{implicit_analytic2}-\ref{implicit_analytic}). For reference it is also displayed in the flow with cooling, using the same adiabatic advection time, which should not be confused with the actual advection time taking into account the cooling layer.
In the adiabatic flow the winding angle of the spiral pattern of advected perturbations from the shock to the neutron star is $\Phi_{\rm adv}\equiv 2\pi-\varphi_{\cal Q}$. We note in Fig.~\ref{fig_phase} that $\varphi_{\cal Q}<\pi$, implying that the radial structure of advected perturbations of entropy or vorticity contains at least a change of sign as seen in Fig.~2 in \cite{Blondin_Shaw2007}, Fig.~1 in \cite{Fernandez2010}, and Fig.~10 in \cite{Buellet2023}.

\subsection{Comparison to the analytic model in \cite{Foglizzo2009}}

The adiabatic formulation incorporates major improvements compared to the adiabatic toy model used by F09; \cite{Sato2009, Guilet2012}: the plane parallel model allowed for explicit analytic solutions and a deeper understanding of the instability mechanism, but some simplifications were difficult to justify: (i) the uniform character of the flow between the shock and the region of deceleration implied that the vertical size of the acoustic and advected cavities were identical, thus overestimating the impact of radial acoustic propagation on the phase of the solution, (ii) the maximum strength ${\cal Q}\sim2$ of the advective-acoustic cycle was arbitrarily set by the specific value ($0.75$) chosen for the ratio $c^2_{\rm sh}/c^2_{\rm out}$. This limited the relative effect of the purely acoustic cycle on the most unstable modes, (iii) the optimum value of the vertical wavenumber compared to the horizontal one was set by the specific value chosen for $L/H=4$ in F09; \cite{Sato2009} and $L/H=6$ in \cite{Guilet2012}, (iv) the width of the coupling region described by $H_\nabla/H$ was a also a free parameter of the model.

The present adiabatic model offers a new framework overcoming these shortcomings, still simple enough to allow for analytical results:
\par-the spherical geometry is taken into account (rather than a plane parallel approximation in F09),
\par-the adiabatic heating and deceleration are produced in a self consistent manner by the radial convergence in 3D and by the proto-neutron star gravity (rather than localized at the lower boundary by some external potential with adjustable depth $c^2_{\rm sh}/c^2_{\rm out}$ and width $H_\nabla/H$ in F09),
\par-the gravity produced by the neutron star at the shock is not neglected, allowing for both displacement and velocity effects for the production of entropy at the shock (rather than ignoring displacement effects with $\omega_\Phi=0$ in F09).

The adiabatic model sheds light on some objections by \cite{Blondin2006} regarding the advective-acoustic mechanism in a non-rotating flow:
\par(i) the fact that the advection time may be longer than the oscillation period does not contradict the advective-acoustic mechanism. 
\par(ii) the fact that the pressure field shows no evidence for a radially localized coupling radius is explained by the radially extended character of the coupling process, as seen in the integral in Eq.~(\ref{single_disp}). The feedback should refer to the radially extended pressure structure rather than the radial propagation of an acoustic wave.

\section{Conclusions and perspectives}
\label{Sect_discussion}

 \begin{enumerate}
 
\item The oscillation period of the fundamental SASI mode can be estimated from the shock radius $r_{\rm sh}$, the radius of maximum deceleration $r_\nabla$ and the central mass $M_{\rm ns}$.

\item A possible improvement to the analytic formula used by \cite{Muller_Janka2014} is proposed, based on the perturbative analysis without any adjustment on a specific numerical simulation.

\item An adiabatic model of the shock dynamics incorporating non adiabatic processes in the boundary conditions is able to capture the general properties of SASI eigenmodes.

\item In the adiabatic approximation the SASI mechanism can be described as a self-forced oscillator. The forcing by advected perturbations is distributed from the shock surface to the cooling layer rather than inside the cooling layer. 

\item An analytic description of the properties of the fundamental mode of SASI in the adiabatic approximation is obtained in the asymptotic limit of a large ratio $r_{\rm sh}/r_\nabla$.

\end{enumerate}

Improving the accuracy of the estimation of the SASI oscillation period from a perturbative analysis requires (i) a more accurate prescription for partial dissociation at the shock and in the flow, as described in Appendix~A in \cite{Fernandez2009b}, and (ii) a better characterization of the parametrized cooling function leading to a realistic value of the ratio $r_\nabla/r_{\rm ns}$.\\

The adiabatic framework opens a new path to investigate analytically the physical effect of stellar rotation in the equatorial plane (paper~II), and address the intriguing results of \cite{Walk2023} obtained by comparing SASI properties in cylindrical and spherical geometries.\\

We note that the adiabatic approximation does account for the growth rate and oscillation period within $\sim30\%$, leaving room for additional non-adiabatic effects. By decreasing the velocity and Mach number near the proto-neutron star they increase the amplitude of the forcing term and also increase the phase mixing near the proto-neutron star. Beside affecting the trade-off between these two effects, non-adiabatic effects modify the amplitude of $\delta S$ and $\delta K$ during their advection, add a source of feedback in the cooling layer and modify the acoustic equation. A quantitative assessment of some of these effects can be obtained within the formalism of the self-forced oscillator, by incorporating them in the lower boundary condition. 

\begin{acknowledgements}
      The anonymous referee is thanked for constructive suggestions.
      It is a pleasure to acknowledge stimulating discussions with J\'er\^ome Guilet, Matteo Bugli, Thomas Janka, Bernhard M\"uller, Kei Kotake, Tomoya Takiwaki, Ernazar Abdikamalov, Laurie Walk, Irene Tamborra, Anne-C\'ecile Buellet, Sonia El Hedri, Florent Robinet and the LEAK collaboration funded by the LabEx UnivEarthS. This work has been financially supported by the PNHE.
\end{acknowledgements}

%
%

\bibliographystyle{aa}
\bibliography{references.bib}

\begin{thebibliography}{38}
\expandafter\ifx\csname natexlab\endcsname\relax\def\natexlab#1{#1}\fi

\bibitem[{{Blondin} {et~al.}(2017){Blondin}, {Gipson}, {Harris}, \&
  {Mezzacappa}}]{Blondin2017}
{Blondin}, J.~M., {Gipson}, E., {Harris}, S., \& {Mezzacappa}, A. 2017, \apj,
  835, 170

\bibitem[{{Blondin} \& {Mezzacappa}(2006)}]{Blondin2006}
{Blondin}, J.~M. \& {Mezzacappa}, A. 2006, \apj, 642, 401

\bibitem[{{Blondin} \& {Mezzacappa}(2007)}]{Blondin_Mezzacappa2007}
{Blondin}, J.~M. \& {Mezzacappa}, A. 2007, \nat, 445, 58

\bibitem[{{Blondin} {et~al.}(2003){Blondin}, {Mezzacappa}, \&
  {DeMarino}}]{Blondin2003}
{Blondin}, J.~M., {Mezzacappa}, A., \& {DeMarino}, C. 2003, \apj, 584, 971

\bibitem[{{Blondin} \& {Shaw}(2007)}]{Blondin_Shaw2007}
{Blondin}, J.~M. \& {Shaw}, S. 2007, \apj, 656, 366

\bibitem[{{Buellet} {et~al.}(2023){Buellet}, {Foglizzo}, {Guilet}, \&
  {Abdikamalov}}]{Buellet2023}
{Buellet}, A.~C., {Foglizzo}, T., {Guilet}, J., \& {Abdikamalov}, E. 2023,
  \aap, 674, A205

\bibitem[{{Burrows} \& {Vartanyan}(2021)}]{Burrows2021}
{Burrows}, A. \& {Vartanyan}, D. 2021, \nat, 589, 29

\bibitem[{{Drago} {et~al.}(2023){Drago}, {Andresen}, {Di Palma}, {Tamborra}, \&
  {Torres-Forn{\'e}}}]{Drago2023}
{Drago}, M., {Andresen}, H., {Di Palma}, I., {Tamborra}, I., \&
  {Torres-Forn{\'e}}, A. 2023, \prd, 108, 103036

\bibitem[{{Dunham} {et~al.}(2024){Dunham}, {Endeve}, {Mezzacappa}, {Blondin},
  {Buffaloe}, \& {Holley-Bockelmann}}]{Dunham2024}
{Dunham}, S.~J., {Endeve}, E., {Mezzacappa}, A., {et~al.} 2024, \apj, 964, 38

\bibitem[{{Fern{\'a}ndez}(2010)}]{Fernandez2010}
{Fern{\'a}ndez}, R. 2010, \apj, 725, 1563

\bibitem[{{Fern{\'a}ndez} {et~al.}(2014){Fern{\'a}ndez}, {M{\"u}ller},
  {Foglizzo}, \& {Janka}}]{Fernandez2014}
{Fern{\'a}ndez}, R., {M{\"u}ller}, B., {Foglizzo}, T., \& {Janka}, H.-T. 2014,
  \mnras, 440, 2763

\bibitem[{{Fern{\'a}ndez} \& {Thompson}(2009{\natexlab{a}})}]{Fernandez2009b}
{Fern{\'a}ndez}, R. \& {Thompson}, C. 2009{\natexlab{a}}, \apj, 703, 1464

\bibitem[{{Fern{\'a}ndez} \& {Thompson}(2009{\natexlab{b}})}]{Fernandez2009a}
{Fern{\'a}ndez}, R. \& {Thompson}, C. 2009{\natexlab{b}}, \apj, 697, 1827

\bibitem[{{Foglizzo}(2001)}]{Foglizzo2001}
{Foglizzo}, T. 2001, \aap, 368, 311

\bibitem[{{Foglizzo}(2002)}]{Foglizzo2002}
{Foglizzo}, T. 2002, \aap, 392, 353

\bibitem[{{Foglizzo}(2009)}]{Foglizzo2009}
{Foglizzo}, T. 2009, \apj, 694, 820

\bibitem[{{Foglizzo} {et~al.}(2005){Foglizzo}, {Galletti}, \&
  {Ruffert}}]{Foglizzo2005}
{Foglizzo}, T., {Galletti}, P., \& {Ruffert}, M. 2005, \aap, 435, 397

\bibitem[{{Foglizzo} {et~al.}(2007){Foglizzo}, {Galletti}, {Scheck}, \&
  {Janka}}]{Foglizzo2007}
{Foglizzo}, T., {Galletti}, P., {Scheck}, L., \& {Janka}, H.~T. 2007, \apj,
  654, 1006

\bibitem[{{Foglizzo} {et~al.}(2015){Foglizzo}, {Kazeroni}, {Guilet}, {Masset},
  {Gonz{\'a}lez}, {Krueger}, {Novak}, {Oertel}, {Margueron}, {Faure}, {Martin},
  {Blottiau}, {Peres}, \& {Durand}}]{Foglizzo2015}
{Foglizzo}, T., {Kazeroni}, R., {Guilet}, J., {et~al.} 2015, \pasa, 32, e009

\bibitem[{{Foglizzo} {et~al.}(2012){Foglizzo}, {Masset}, {Guilet}, \&
  {Durand}}]{Foglizzo2012}
{Foglizzo}, T., {Masset}, F., {Guilet}, J., \& {Durand}, G. 2012, \prl, 108,
  051103

\bibitem[{{Foglizzo} {et~al.}(2006){Foglizzo}, {Scheck}, \&
  {Janka}}]{Foglizzo2006}
{Foglizzo}, T., {Scheck}, L., \& {Janka}, H.~T. 2006, \apj, 652, 1436

\bibitem[{{Guilet} \& {Foglizzo}(2012)}]{Guilet2012}
{Guilet}, J. \& {Foglizzo}, T. 2012, \mnras, 421, 546

\bibitem[{{Huete} {et~al.}(2018){Huete}, {Abdikamalov}, \&
  {Radice}}]{Huete2018}
{Huete}, C., {Abdikamalov}, E., \& {Radice}, D. 2018, \mnras, 475, 3305

\bibitem[{{Janka}(2017)}]{Janka2017}
{Janka}, H.-T. 2017, \apj, 837, 84

\bibitem[{{Janka} {et~al.}(2016){Janka}, {Melson}, \& {Summa}}]{Janka2016}
{Janka}, H.-T., {Melson}, T., \& {Summa}, A. 2016, Annual Review of Nuclear and
  Particle Science, 66, 341

\bibitem[{{Kitaura} {et~al.}(2006){Kitaura}, {Janka}, \&
  {Hillebrandt}}]{Kitaura2006}
{Kitaura}, F.~S., {Janka}, H.~T., \& {Hillebrandt}, W. 2006, \aap, 450, 345

\bibitem[{{Kotake} \& {Kuroda}(2017)}]{Kotake2017}
{Kotake}, K. \& {Kuroda}, T. 2017, in Handbook of Supernovae, ed. A.~W.
  {Alsabti} \& P.~{Murdin}, 1671

\bibitem[{{M{\"u}ller}(2019{\natexlab{a}})}]{Muller2019b}
{M{\"u}ller}, B. 2019{\natexlab{a}}, \mnras, 487, 5304

\bibitem[{{M{\"u}ller}(2019{\natexlab{b}})}]{Muller2019a}
{M{\"u}ller}, B. 2019{\natexlab{b}}, Annual Review of Nuclear and Particle
  Science, 69, 253

\bibitem[{{M{\"u}ller}(2020)}]{Muller2020}
{M{\"u}ller}, B. 2020, Living Reviews in Computational Astrophysics, 6, 3

\bibitem[{{M{\"u}ller} \& {Janka}(2014)}]{Muller_Janka2014}
{M{\"u}ller}, B. \& {Janka}, H.-T. 2014, \apj, 788, 82

\bibitem[{{M{\"u}ller} {et~al.}(2019){M{\"u}ller}, {Tauris}, {Heger},
  {Banerjee}, {Qian}, {Powell}, {Chan}, {Gay}, \& {Langer}}]{Muller_Tauris2019}
{M{\"u}ller}, B., {Tauris}, T.~M., {Heger}, A., {et~al.} 2019, \mnras, 484,
  3307

\bibitem[{{Powell} \& {M{\"u}ller}(2022)}]{Powell2022}
{Powell}, J. \& {M{\"u}ller}, B. 2022, \prd, 105, 063018

\bibitem[{{Sato} {et~al.}(2009){Sato}, {Foglizzo}, \& {Fromang}}]{Sato2009}
{Sato}, J., {Foglizzo}, T., \& {Fromang}, S. 2009, \apj, 694, 833

\bibitem[{{Scheck} {et~al.}(2008){Scheck}, {Janka}, {Foglizzo}, \&
  {Kifonidis}}]{Scheck2008}
{Scheck}, L., {Janka}, H.~T., {Foglizzo}, T., \& {Kifonidis}, K. 2008, \aap,
  477, 931

\bibitem[{{Stockinger} {et~al.}(2020){Stockinger}, {Janka}, {Kresse}, {Melson},
  {Ertl}, {Gabler}, {Gessner}, {Wongwathanarat}, {Tolstov}, {Leung}, {Nomoto},
  \& {Heger}}]{Stockinger2020}
{Stockinger}, G., {Janka}, H.~T., {Kresse}, D., {et~al.} 2020, \mnras, 496,
  2039

\bibitem[{{Tamborra} \& {Murase}(2019)}]{Tamborra2019}
{Tamborra}, I. \& {Murase}, K. 2019, in Supernovae. Series: Space Sciences
  Series of ISSI, ed. A.~{Bykov}, C.~{Roger}, J.~{Raymond}, F.-K. {Thielemann},
  M.~{Falanga}, \& R.~{von Steiger}, Vol.~68, 87--107

\bibitem[{{Walk} {et~al.}(2023){Walk}, {Foglizzo}, \& {Tamborra}}]{Walk2023}
{Walk}, L., {Foglizzo}, T., \& {Tamborra}, I. 2023, \prd, 107, 063014

\end{thebibliography}

\begin{appendix}


\section{Stationary accretion}\label{sec:Adiff}

The loss of energy through nuclear dissociation is measured by the parameter $\varepsilon$ defined by Eq.~(\ref{norm_disso}). It decreases the value of the post-shock Mach number $\M_{\rm sh}$ below the reference adiabatic value $\M_{\rm ad}$, and affects the Rankine-Hugoniot conditions according to Eqs.~(A4-A6) of \cite{Foglizzo2006}:
\begin{eqnarray}
\M_{\rm ad}^2\equiv{2+(\gamma-1)\M_1^2\over 2\gamma\M_1^2-\gamma+1},\label{def_Mad}\\
\varepsilon=\left(1+{2\over\gamma-1}{1\over\M_1^2}\right)
\left(1-{\M_{\rm sh}^2\over\M_{\rm ad}^2}\right)\left(1-{\M_{\rm sh}^2\over\M_{1}^2}\right),
\label{varepsilon}\\
{\v2\over v_1}={\M_{\rm sh}^2\over\M_{1}^2}{1+\gamma\M_{1}^2\over 1+\gamma\M_{\rm sh}^2},\label{v2v1}\\
{c_{\rm sh}\over c_1}={\M_{\rm sh}\over\M_{1}}{1+\gamma\M_{1}^2\over 1+\gamma\M_{\rm sh}^2}.\label{c2c1}
\end{eqnarray}
The nuclear binding energy of the iron nuclei is $8.8$MeV per nucleon, or $1.77\times10^{52}{\rm erg}/M_\odot$. If nuclear dissociation was complete across the shock this would translates into a dissociation parameter scaling linearly with $r_{\rm sh}$:
\begin{eqnarray}
\varepsilon&=&0.7\left({r_{\rm sh}\over 150{\rm km}}\right)\left({1.3M_\odot\over M_{\rm ns}}\right).
\label{epsilon_r}
\end{eqnarray}
We note that our definition of $\varepsilon$ in Eq.~(\ref{norm_disso}) follows \cite{Huete2018}, which differs from Eq.~(4) in \cite{Fernandez2009a} by a factor $2$. A fraction of the dissociation of iron nuclei is radially distributed between the shock and the neutron star surface, as calculated in \cite{Fernandez2009b}. Taking into account incomplete dissociation, the value of $\epsilon$ formulated in Eq.~(\ref{disso0.5}) is a rough estimate deduced from Eq.~(51) and Fig.~12 in \cite{Huete2018} for $r_{\rm sh}<175$km. It is smaller than Eq.~(\ref{epsilon_r}) by a factor $\sim1.4$ and saturates at $\epsilon\sim 0.5$ in exploding models.\\
The value of $\M_{\rm sh}$ is expressed as a function of $\M_1$ and $\varepsilon(r_{\rm sh})$ using Eq.~(\ref{varepsilon}):
\begin{eqnarray}
{\M_{\rm sh}^2\over\M_{\rm ad}^2}
&=&
{1\over2}
+
{
1-{2\varepsilon\over1+{2\over\gamma-1}{1\over\M_1^2}}-{\M_{\rm ad}^2\over2\M_1^2}
\over
1+\left\lbrack
\left(1-{\M_{\rm ad}^2\over\M_1^2}\right)^2+{4\varepsilon{\M_{\rm ad}^2\over\M_1^2}\over1+{2\over\gamma-1}{1\over\M_1^2}}
\right\rbrack^{1\over2}
}
\label{Msh_epsilon}
,\\
&\sim& 1-\varepsilon\;\;{\rm if}\;\;\M_1\gg1
\end{eqnarray}
The jump condition (\ref{v2v1}) is thus a function of $\varepsilon(r_{\rm sh})$ using Eq.~(\ref{Msh_epsilon}), with the simple following expression for a strong shock:
\begin{eqnarray}
{v_1\over \v2}&\sim& 1+ {2\over (1-\varepsilon)(\gamma-1)},
\label{v2v1_strong}\\
&\sim& 1+ {6\over 1-\varepsilon}.
\end{eqnarray}
The definition of the cooling function (\ref{def_cooling}), the dimensionless entropy (\ref{def_entropy}) and the mass conservation (\ref{eq_mass_stat}) are used to eliminate the variables $\rho,p,\M,v$ from the stationary equations (\ref{eq_entropy_stat}-\ref{eq_euler_stat}):
\begin{eqnarray}
{\rho\over\rho_{\rm sh}}&=&\left({c\over c_{\rm sh}}\right)^{2\over\gamma-1}{\rm e}^{-(S-S_{\rm sh})},\\
r^gc^{\gamma+1\over\gamma-1}\M&=&r_{\rm sh}^gc_{\rm sh}^{\gamma+1\over\gamma-1}{\rm e}^{S-S_{\rm sh}}  .\label{mass_cons}
\end{eqnarray}

\section{Differential system and boundary conditions of the perturbed accretion}
\label{append_nonax}

We rewrite the differential system satisfied by $\delta f$, $\delta h$ (Eqs.~(E2-E3) in \cite{Foglizzo2006}), using the radial coordinate $X$ defined as in \cite{Foglizzo2001}:
\begin{eqnarray}
{\rm d X}&\equiv& {v \over 1-\M^2} {\rm d}r.
\end{eqnarray} 
Thus
\begin{eqnarray}
\left({\partial \over\partial X}+{i\omega\over c^2}\right)
{\delta f\over i\omega}=
\delta h
+\left(\gamma-1+{1\over\M^2}\right){\delta S\over\gamma}\nonumber\\
+{1-\M^2\over i\omega v}\delta\left({{\cal L}\over\rho v}\right),\label{dfdX} \\
\left({\partial \over\partial X}+{i\omega\over c^2}\right) \delta h
={i\omega\over v^2}\left({\omega^2-\omega_{\rm Lamb}^2\over \omega^2c^2}\delta f-\delta S\right)
+{1-\M^2\over v^2}{i\delta K\over\omega r^2}
\label{dhdX}
\end{eqnarray} 
The set of equations (A1-A4) in \cite{Foglizzo2007} are repeated here for completeness:
\begin{eqnarray}
 {\delta v_r \over v }&=&{1\over 1-\M^2}\left(\delta h+\delta S-{\delta f\over c^2} \right),\label{dvr}\\
{\delta \rho\over\rho}&=&{1\over 1-\M^2}\left(-\M^2\delta h-\delta S+{\delta f\over c^2} \right),\label{drho}\\
{\delta c^2\over c^2}&=&{\gamma-1\over 1-\M^2}\left(
{\delta f\over c^2}-\M^2\delta h-\M^2\delta S \right),\label{dc2}\\
{\delta p\over \gamma p}&=&{1\over\gamma-1}{\delta c^2\over c^2}
-{\delta S\over\gamma} ,\label{dP}\\
\delta \left( {{\cal L}\over\rho v} \right) &=& \nabla S {c^2\over\gamma} \left\lbrack (\beta - 1) {\delta \rho\over\rho} + \alpha {\delta c^2\over c^2} - {\delta v_r\over v} \right\rbrack\ ,
\\
\delta \left( {{\cal L}\over p v} \right) &=& 
\nabla S \left\lbrack (\beta - 1) {\delta \rho\over\rho} + (\alpha-1) {\delta c^2\over c^2} - {\delta v_r\over v} \right\rbrack .
\end{eqnarray}
The perturbed mass conservation and transverse components of the perturbed Euler equation are:
\begin{eqnarray}
-i\omega\delta v_\theta+v\delta w_\varphi +{1\over r}{\partial\over\partial \theta}\delta f=
{c^2\over\gamma}{1\over r}{\partial\delta S\over\partial \theta}
,
\label{ddfdtheta}
\\
-i\omega\delta v_\varphi-v\delta w_\theta +{im\over r\sin\theta}\delta f=
{c^2\over\gamma}{im\delta S\over r\sin\theta}
,
\label{ddfdphi}
\end{eqnarray}
We eliminate $\delta v_\theta$ and $\delta v_\varphi$ in the system of Eqs.~(\ref{defdK0}), (\ref{def_dA}), (\ref{ddfdtheta}) and (\ref{ddfdphi}) and obtain Eq.~(\ref{AFK}).\\
The derivative of $\delta A$ is calculated using Eqs.~(\ref{AFK}), (\ref{dKdr_main}) and (\ref{dfdX}), leading to the differential system (\ref{dAdr_main}-\ref{dKdr_main}) satisfied by $\delta A,\delta h,\delta S,\delta K$.
The second derivative of $\delta A$ is calculated using Eqs.~(\ref{defdK0}), (\ref{dKdr_main}) and (\ref{dhdX}), resulting in Eq.~(\ref{eqdiffAcool_main}).\\
Using Eq.~(\ref{AFK}), Eqs.~(\ref{dvr}-\ref{dP}) are rewritten as follows:
\begin{eqnarray}
 {\delta v_r \over v }&=&
 {\delta K\over \ell(\ell+1)v^2} 
-
{1\over \ell(\ell+1)v}{\partial \delta A\over\partial r}
-
{\delta S\over\gamma\M^2}
 ,\label{dvrA}\\
{\delta \rho\over\rho}
&=&
{1\over c^2}
 \left( v{\partial \over\partial r} - i\omega\right)
 {\delta A\over \ell(\ell+1)}
-{\gamma-1\over\gamma}\delta S
,\label{drhoA}\\
{1\over\gamma-1}{\delta c^2\over c^2}&=&
{1\over c^2}
 \left( v{\partial \over\partial r} - i\omega\right)
 {\delta A\over \ell(\ell+1)}
+
{\delta S\over\gamma}
 ,\label{dc2A}\\
 {\delta p\over \gamma p}&=&{1\over c^2}
 \left( v{\partial \over\partial r} - i\omega\right)
 {\delta A\over \ell(\ell+1)}
 ,\label{dPA}
\end{eqnarray}
Using Eq.~(\ref{dvrA}) with Eq.~(\ref{defdK0}), we can express $\delta v_r$ with $\delta A$ and $\delta w_\perp$ and obtain Eq.~(\ref{dvrw_main}). We note that this relation is equivalent to the radial component of the vector calculus relation $\nabla^2v=\nabla(\nabla\cdot v)-\nabla\times(\nabla\times v)$. \\
The baroclinic production of vorticity by the advection of entropy perturbations can be calculated in the same way as Eqs.~(E17-E19) in \cite{Foglizzo2005}, using the conservation of the tangential component of the velocity across the shock:
\begin{eqnarray}
\delta v_{\theta{\rm sh}}&=&{v_1-v_{\rm sh}\over r_{\rm sh}}{\partial\Delta \zeta\over\partial\theta},\label{dvtheta}\\
\delta v_{\varphi{\rm sh}}&=&{v_1-v_{\rm sh}\over r_{\rm sh}}{im\Delta \zeta\over \sin\theta}.\label{dvphi}
\end{eqnarray} 
The vorticity produced at the shock is defined by Eqs.~(\ref{dvtheta}-\ref{dvphi}), the transverse components (\ref{ddfdtheta}-\ref{ddfdphi}) of the Euler equation at the shock, together with the angular derivative of Eqs.~(\ref{dAsh}):
\begin{eqnarray}
\delta w_{r{\rm sh}} &=&0,\label{dw_r}\\
\delta w_{\theta{\rm sh}} &=&
-
{c_{\rm sh}^2\over\gamma}{im\over r_{\rm sh}v_{\rm sh}\sin\theta}
\left(\delta S_{\rm sh}+\left\lbrack\nabla S\right\rbrack^{\rm sh}_{1}
\Delta\zeta\right)
,
\label{dw_theta}
\\
\delta w_{\varphi{\rm sh}} &=&
{c_{\rm sh}^2\over\gamma}
{1\over r_{\rm sh}v_{\rm sh}}
{\partial\over\partial \theta}\left(\delta S_{\rm sh}+\left\lbrack\nabla S\right\rbrack^{\rm sh}_{1}
\Delta\zeta\right)
.
\label{dw_phi}
\end{eqnarray}

\section{Adiabatic model}
\label{App_adiab}

Following Eqs.~(B5-B7) in \cite{Foglizzo2001}, the differential equation describing the specific vorticity in a spherical adiabatic flow can be integrated as
\begin{eqnarray}
\delta w_r&=&\left({r_{\rm sh}\over r}\right)^2\delta w_{r{\rm sh}}{\rm e}^{i\omega\int_{\rm sh}{{\rm d}r\over v}},
\label{dwr_F01}\\
\delta w_\theta&=&{1\over rv}\left\lbrack (rv\delta w_\theta)_{\rm sh}
-{c^2-c_{\rm sh}^2\over\sin\theta}im{\delta S_{\rm sh}\over\gamma}\right\rbrack{\rm e}^{i\omega\int_{\rm sh}{{\rm d}r\over v}}
,\label{dwtheta_F01}\\
\delta w_\varphi&=&{1\over rv}\left\lbrack (rv\delta w_\varphi)_{\rm sh}
+{c^2-c_{\rm sh}^2\over\sin\theta}{\partial\over\partial\theta}{\delta S_{\rm sh}\over\gamma}\right\rbrack{\rm e}^{i\omega\int_{\rm sh}{{\rm d}r\over v}}.\label{dwphi_F01}
\end{eqnarray}
Using Eqs.~(\ref{dw_r}-\ref{dw_phi}) with $\nabla S=0$, together with Eqs.~(\ref{dwr_F01}-\ref{dwphi_F01}) gives the expression (\ref{dwr_main}-\ref{dwphi_main}) of the vorticity perturbation throughout the flow.\\
With $\delta Y$ defined by Eq.~(\ref{def_Y}), the expressions of $\delta A_{\rm sh}$, $\delta h_{\rm sh}$ are deduced from Eqs.~(\ref{dAsh}-\ref{dSsh_zeta}):
\begin{eqnarray}
\left(1-{\v2\over v_{1}}\right)\Delta\zeta={\delta Y_{\rm sh}\over v_1},\\
\delta h_{\rm sh}=-{i\omega \over v_1v_{\rm sh}}
\delta Y_{\rm sh},\\
{\delta S_{\rm sh}\over\gamma} = 
  {\delta Y_{\rm sh} \over c_{\rm sh}^2}
(i\omega+\omega_\Phi) \left(1-{\v2\over v_{1}}\right).
\end{eqnarray} 
We rewrite Eq.~(\ref{dAdr_main}) using the definitions of $X$ and $\delta Y$ with $\delta K=0$:
\begin{eqnarray}
\delta Y_{\rm sh}=-{\delta A_{\rm sh}\over \ell(\ell+1)},\\
\left({\partial \delta A\over\partial r}\right)_{\rm sh}
={\ell(\ell+1)v_{\rm sh}\over 1-\M_{\rm sh}^2}\left\lbrack
\delta Y_{\rm sh}{i\omega  \over v_{\rm sh}^2}\left({\v2\over v_1}+\M_{\rm sh}^2\right)
\right.\nonumber\\
\left.
-{\delta S_{\rm sh}\over\gamma }\left({1\over\M_{\rm sh}^2}+\gamma-1\right)\right\rbrack,\\
\left({\partial\delta Y\over\partial X}\right)_{\rm sh}
=
-{1-\M_{\rm sh}^2\over \ell(\ell+1)v_{\rm sh}}\left({\partial 
\delta A\over\partial r}\right)_{\rm sh}+{i\omega\over c_{\rm sh}^2}\delta Y_{\rm sh} ,
\end{eqnarray} 
which results in Eq.~(\ref{dYdXsh_0}).\\
At the inner boundary in the adiabatic approximation, we use the condition $\delta v_r=0$:
\begin{eqnarray}
\delta h_{\rm ns}+\delta S_{\rm ns}={\delta f\over   c_{\rm ns}^2} .\label{innerbound}
\end{eqnarray}
The entropy is simply advected from the shock (\ref{integ_dS}) and we express $\delta h$ with $\delta Y$ and $\partial\delta Y/\partial X$ using Eq.~(\ref{dAdr_main}):
\begin{eqnarray}
\delta S_{\rm ns}&=&\delta S_{\rm sh}{\rm e}^{\int_{\rm sh}^{\rm ns}i\omega {{\rm d}r\over v}},\\
{\partial\delta Y\over\partial X}&=&\delta h{\rm e}^{\int_{\rm sh}{i\omega\over c^2}{\rm d}X}
\nonumber\\
&&+{\delta S\over\gamma}{\rm e}^{\int_{\rm sh}{i\omega\over c^2}{\rm d}X}\left({1\over\M^2}+\gamma-1\right).
\end{eqnarray}
The inner boundary condition (\ref{innerbound}) is thus reduced to Eq.~(\ref{lowerbc_0}).\\
Multiplying  Eq.~(\ref{forced_oscillator}) by $m/(\omega\sin\theta)$ and using Eqs.~(\ref{dwtheta_main}) and (\ref{rdvphi}):
\begin{eqnarray}
\left\lbrack\left({\partial\over\partial X}+{i\omega\over c^2}\right)^2
+
{\omega^2-\omega_{\rm Lamb}^2\over v^2c^2}\right\rbrack
r\delta v_\varphi
=-{\partial\over\partial X}
{r\delta w_\theta\over v}.\label{diffspheriquephi}
\end{eqnarray} 
Equivalently, using Eq.~(\ref{dwphi_main}) and (\ref{rdvtheta}) and the derivative of Eq.~(\ref{forced_oscillator}) with respect to $\theta$ leads to:
\begin{eqnarray}
\left\lbrack\left({\partial\over\partial X}+{i\omega\over c^2}\right)^2
+
{\omega^2-\omega_{\rm Lamb}^2\over v^2c^2}\right\rbrack
r\delta v_\theta
={\partial\over\partial X}
{r\delta w_\varphi\over v},\label{diffspheriquetheta}
\end{eqnarray} 
which is equivalent to Eq.~(\ref{diffspheriquephi}) given the spherical symmetry of the stationary flow.

\section{Equation defining the eigenfrequencies in the adiabatic model}
\label{integral_dispersion}

The differential equation satisfied by $\delta Y$ is deduced from Eqs.~(\ref{forced_oscillator}) and (\ref{def_Y}):
\begin{eqnarray}
\left({\partial^2\over\partial X^2}
+
{\omega^2-\omega_{\rm Lamb}^2\over v^2c^2}\right)
\delta Y
=
\delta{\cal F},\label{forced_dY}\\
\delta{\cal F}\equiv
{\rm e}^{\int_{\rm sh} {i\omega\over c^2}{\rm d}X}{\cal F}_S\delta S_{\rm sh}.
\label{def_dF}
\end{eqnarray} 
We define $Y_0$ as the solution of the homogeneous equation satisfying the inner boundary condition of pure acoustic waves (i.e. without entropy and vorticity perturbations), and $Y_-$ another homogeneous solution such that their Wronskien is $W$:
\begin{eqnarray}
\left({\partial Y_0\over\partial X}\right)_{\rm ns}={i\omega \over   c_{\rm ns}^2} Y_0^{\rm ns},\label{lowerbc0}\\
W\equiv Y_0{\partial Y_-\over\partial X}-Y_-{\partial Y_0\over\partial X},\label{Wronskien}\\
\delta Y=Y_-\left(d_-+\int_{\rm ns}{ Y_0\over W} \delta{\cal F}{\rm d}X\right)
-Y_0\left(d_0+\int_{\rm sh}{ Y_-\over W} \delta{\cal F}{\rm d}X\right),
\label{def_Y_Green}
\end{eqnarray}
where $\delta{\cal F}\equiv {\cal F}_S\delta S_{\rm sh}$ is the forcing term on the right hand side of Eq.~(\ref{forced_oscillator}) for the variable $\delta Y$ defined by .
Using Eq.~(\ref{def_Y_Green}) at the upper boundary:
\begin{eqnarray}
\delta Y_{\rm sh}=Y_-^{\rm sh}\left(d_-+\int_{\rm ns}^{\rm sh}{ Y_0\over W} \delta{\cal F}{\rm d}X\right)
-d_0 Y_0^{\rm sh},\label{Ysh}\\
\left({\partial\delta Y\over\partial X}\right)_{\rm sh}=
\left({\partial Y_-\over\partial X}\right)_{\rm sh}\left(d_-+\int_{\rm ns}^{\rm sh}{ Y_0\over W} \delta{\cal F}{\rm d}X\right)
-d_0 \left({\partial Y_0\over\partial X}\right)_{\rm sh},\label{dYsh}
\end{eqnarray}
Using Eq.~(\ref{def_Y_Green}) at the lower boundary:
\begin{eqnarray}
\delta Y_{\rm ns}=d_-Y_-^{\rm ns}-Y_0^{\rm ns}\left(d_0-\int_{\rm ns}^{\rm sh}{ Y_-\over W} \delta{\cal F}{\rm d}X\right),\\
\left({\partial\delta Y\over\partial X}\right)_{\rm ns}=
d_-\left({\partial Y_-\over\partial X}\right)_{\rm ns}
-\left({\partial Y_0\over\partial X}\right)_{\rm ns}\left(d_0-\int_{\rm ns}^{\rm sh}{ Y_-\over W} \delta{\cal F}{\rm d}X\right).
\end{eqnarray}
Using Eq.~(\ref{lowerbc0}), the lower boundary condition (\ref{lowerbc_0}) translates into
\begin{eqnarray}
d_-\left\lbrack \left({\partial Y_-\over\partial X}\right)_{\rm ns} - {i\omega \over   c_{\rm ns}^2} Y_-^{\rm ns}\right\rbrack
=
\delta{\cal F}_{\rm sh}(1-\M^2_{\rm ns}){\M_{\rm sh}^2\over\M^2_{\rm ns}}{\rm e}^{\int_{\rm sh}^{\rm ns} {i\omega\over v^2}{\rm d}X}.
\end{eqnarray}
We note using Eq.~(\ref{Wronskien}) and (\ref{lowerbc0}) that
\begin{eqnarray}
 \left({\partial Y_-\over\partial X}\right)_{\rm ns} - {i\omega \over   c_{\rm ns}^2} Y_-^{\rm ns}
  &=&{W\over Y_0^{\rm ns}}.
\end{eqnarray}
Thus
\begin{eqnarray}
Wd_-
=
Y_0^{\rm ns}
\delta{\cal F}_{\rm sh}(1-\M^2_{\rm ns}){\M_{\rm sh}^2\over\M^2_{\rm ns}}{\rm e}^{\int_{\rm sh}^{\rm ns}{i\omega\over v^2}{\rm d}X}.
\end{eqnarray}
Eliminating $d_0$ between Eqs.~(\ref{Ysh}) and (\ref{dYsh}) and using Eq.~(\ref{Wronskien}):
\begin{eqnarray}
Y_0^{\rm sh}\left({\partial\delta Y\over\partial X}\right)_{\rm sh}-\left({\partial Y_0\over\partial X}\right)_{\rm sh}
\delta Y_{\rm sh}
=\nonumber\\
\left\lbrack Y_0^{\rm sh}\left({\partial Y_-\over\partial X}\right)_{\rm sh}-\left({\partial Y_0\over\partial X}\right)_{\rm sh}Y_-^{\rm sh}\right\rbrack
\left(d_-+\int_{\rm ns}^{\rm sh}{ Y_0\over W} \delta{\cal F}{\rm d}X\right),\\
=
Wd_-+\int_{\rm ns}^{\rm sh} Y_0 \delta{\cal F}{\rm d}X.
\end{eqnarray}
The eigenfrequencies are thus defined by
\begin{eqnarray}
Y_0^{\rm sh}\left({\partial\delta Y\over\partial X}\right)_{\rm sh}-\left({\partial Y_0\over\partial X}\right)_{\rm sh}
\delta Y_{\rm sh}
=\nonumber\\
Y_0^{\rm ns}
\delta{\cal F}_{\rm sh}(1-\M^2_{\rm ns}){\M_{\rm sh}^2\over\M^2_{\rm ns}}{\rm e}^{\int_{\rm sh}^{\rm ns}{i\omega\over v^2}{\rm d}X}
+\int_{\rm ns}^{\rm sh} Y_0 \delta{\cal F}{\rm d}X.
\end{eqnarray}
Replacing the forcing term by its expression (\ref{def_dF}),
\begin{eqnarray}
Y_0^{\rm sh}\left({\partial\delta Y\over\partial X}\right)_{\rm sh}-\left({\partial Y_0\over\partial X}\right)_{\rm sh}
\delta Y_{\rm sh}
=\nonumber\\
\delta{\cal F}_{\rm sh}\left\lbrace
Y_0^{\rm ns}
(1-\M^2_{\rm ns}){\M_{\rm sh}^2\over\M^2_{\rm ns}}{\rm e}^{\int_{\rm sh}^{\rm ns}{i\omega\over v^2}{\rm d}X}
\right.\nonumber\\
\left.
+\int_{\rm ns}^{\rm sh} Y_0 
{\rm e}^{\int_{\rm sh} {i\omega\over c^2}{\rm d}X}
{\partial \over\partial r}
\left(
{\M_{\rm sh}^2\over\M^2}
{\rm e}^{\int_{\rm sh} {i\omega\over v}{\rm d}r}
\right){\rm d}r
\right\rbrace.
\end{eqnarray}
Using Eqs.~(\ref{dSsh_zeta}) and (\ref{dYdXsh_0}) this equation takes the form
\begin{eqnarray}
a_1Y_0^{\rm sh}
+a_2r_{\rm sh}\left({\partial Y_0\over\partial r}\right)_{\rm sh}
+a_3Y_0^{\rm ns}
=\nonumber\\
\int_{\rm ns}^{\rm sh} Y_0 
{\rm e}^{\int_{\rm sh} {i\omega\over c^2}{\rm d}X}
{\partial \over\partial r}
\left(
{\M_{\rm sh}^2\over\M^2}
{\rm e}^{\int_{\rm sh} {i\omega\over v}{\rm d}r}
\right){\rm d}r,\label{dispersion0}
\end{eqnarray}
with $a_1,a_2,a_3$ defined by:
\begin{eqnarray}
a_1&\equiv&
{\left({\partial\delta Y\over\partial X}\right)_{\rm sh}\over \delta{\cal F}_{\rm sh}},\\
&=&1+(\gamma-1)\M_{\rm sh}^2
-
{i\omega  \over i\omega+\omega_\Phi}
{{\v2\over v_1}
\over
1-{\v2\over v_{1}}
}
,\label{defa1}
\\
a_2&\equiv&
-{1-\M_{\rm sh}^2\over r_{\rm sh}v_{\rm sh}}{\delta Y_{\rm sh}\over  \delta{\cal F}_{\rm sh}}
,\\
&=&
-
{
1-\M_{\rm sh}^2
\over
\left(1-{\v2\over v_{1}}\right)
{(i\omega+\omega_\Phi) r_{\rm sh}\over \v2}
}\label{defa2}
,\\
a_3&\equiv&
-(1-\M^2_{\rm ns}){\M_{\rm sh}^2\over\M^2_{\rm ns}}{\rm e}^{\int_{\rm sh}^{\rm ns}  {i\omega\over v^2}{\rm d}X}
.\label{defa3}
\end{eqnarray}
After one integration by parts:
\begin{eqnarray}
Y_0^{\rm sh}\left({\partial\delta Y\over\partial X}\right)_{\rm sh}-\left({\partial Y_0\over\partial X}\right)_{\rm sh}
\delta Y_{\rm sh}
=
\delta{\cal F}_{\rm sh}\left\lbrace
Y_0^{\rm sh} 
-Y_0^{\rm ns}
\M_{\rm sh}^2{\rm e}^{\int_{\rm sh}^{\rm ns}{i\omega\over v^2}{\rm d}X}
\right.\nonumber\\
\left.
-\int_{\rm ns}^{\rm sh} 
{\partial \over\partial r}\left(
Y_0 
{\rm e}^{\int_{\rm sh} {i\omega\over c^2}{\rm d}X}
\right)
{\M_{\rm sh}^2\over\M^2}
{\rm e}^{\int_{\rm sh} {i\omega\over v}{\rm d}r}
{\rm d}r
\right\rbrace
.
\end{eqnarray}
The equation defining the eigenfrequencies becomes Eq.~(\ref{single_disp}) with $a'_1\equiv a_1-1$ and $a'_2\equiv a_2$ defined by Eqs.~(\ref{def_a1p}-\ref{def_a2p}).

\section{Approximation of the adiabatic stationary flow and the homogeneous perturbative solution 
}
\label{app_approx}
The dissociation measured by the parameter $\varepsilon$ affects the relation between $|v_{\rm sh}|$ and the local free fall velocity, as deduced from Eq.~(\ref{v1v2simpl_m}) for a strong shock:
\begin{eqnarray}
|v_{\rm sh}|\left({r_{\rm sh}\over2GM_{\rm ns}}\right)^{1\over2}&\sim&
{(\gamma-1)(1-\varepsilon) \over 2+(\gamma -1)(1-\varepsilon)},\label{vvff}\\
{c_{\rm sh}^2r_{\rm sh}\over GM_{\rm ns}}&=&{1\over \M_{\rm sh}^2}{v_{\rm sh}^2r_{\rm sh}\over GM_{\rm ns}}\label{c2GM}.
\end{eqnarray}  
Equation (\ref{c2GM}) is transformed into Eq.~(\ref{c2G_M_m}) using Eqs.~(\ref{vvff}) and (\ref{Machsimpl_m}).
We use a power law approximation of the enthalpy profile deduced from the Bernoulli equation for $r\ll r_{\rm sh}$ and $\M^2\ll1$:
\begin{eqnarray}
c^2\left\lbrack1+(\gamma-1){\M^2\over2}\right\rbrack=(\gamma-1){GM_{\rm ns}\over r}\nonumber\\
+c_{\rm sh}^2\left\lbrack1+(\gamma-1){\M_{\rm sh}^2\over2}\right\rbrack
-(\gamma-1){GM_{\rm ns}\over r_{\rm sh}},\\
c^2\sim (\gamma-1){GM_{\rm ns}\over r}.\label{Bersimpl}
\end{eqnarray} 
The mass conservation and the adiabatic hypothesis in Eq.~(\ref{mass_cons}) imply the power law approximations (\ref{approx_v}) and (\ref{approx_M}) for the velocity and Mach number profiles. \\
For $\gamma=4/3$ we approximate 
\begin{eqnarray}
\M^2\sim \M_{\rm sh}^2\left({r\over r_{\rm sh}}\right)^3,\\
v\sim v_{\rm sh}{r\over r_{\rm sh}},\\
i\omega\int_{\rm sh} {\M^2\over 1-\M^2}{{\rm d}r\over v}
\sim 
-\M_{\rm sh}^2{i\omega r_{\rm sh}\over |v_{\rm sh}|}\int_{\rm sh} {x^2{\rm d}x\over 1-\M_{\rm sh}^2x^3},\\
\sim
{i\omega r_{\rm sh}\over 3|v_{\rm sh}|} 
\log\left( {1-\M^2\over 1-\M_{\rm sh}^2}\right).
\end{eqnarray}
The oscillatory phase associated to $\delta f$ in Eq.~(\ref{single_disp}) is made of a product of ($i\omega$) with two explicit contributions and a third contribution from the definition of $\delta Y$ in Eq.~(\ref{def_Y}). The sum of these three contributions is:
\begin{eqnarray}
\int_{\rm sh} {\M^2\over 1-\M^2}{{\rm d}r\over v}
+\int_{\rm sh} {\M^2\over 1-\M^2}{{\rm d}r\over v}
+\int_{\rm sh} {{\rm d}r\over v}
\nonumber\\
=\int_{\rm sh} {1+\M^2\over 1-\M^2}{{\rm d}r\over v}.
\end{eqnarray} 
With $\alpha_v=1$ and $\tilde r\equiv r/r_{\rm sh}$,
\begin{eqnarray}
{1+\M^2\over 1-\M^2}{v_{\rm sh}\over v}\sim
{2\M_{\rm sh}^2{\tilde r}^2\over 1-\M_{\rm sh}^2{\tilde r}^3}+{1\over {\tilde r}}.
\end{eqnarray} 
The integral can be estimated as follows:
\begin{eqnarray}
\int_{\rm sh} {1+\M^2\over 1-\M^2}{{\rm d}r\over v}&\sim&
{r_{\rm sh}\over |v_{\rm sh}|}
\left(
\log {r_{\rm sh}\over r_{\rm ns}}
-{2\over 3}\log {1-\M_{\rm sh}^2\over 1-\M_{\rm ns}^2}
\right),\\
&\sim &
{r_{\rm sh}\over |v_{\rm sh}|}
\left(
\log {r_{\rm sh}\over r_{\rm ns}}
+{2\M_{\rm sh}^2\over 3}
\right)
\end{eqnarray} 
With a strong adiabatic shock $\M_{\rm sh}^2\sim 1/8$, the correction to the advection timescale $\tau_{\rm adv}^{\rm ns}$ is thus of order $12\%$ for $r_{\rm sh}/ r_{\rm ns}= 2$ and $5\%$ for $r_{\rm sh}/ r_{\rm ns}= 5$. This correction is smaller if dissociation diminishes the value of $\M_{\rm sh}^2$.\\
The contribution of the integral inside the radial derivative in Eq.~(\ref{single_disp}) is limited to a contribution of order $\M_{\rm sh}^2$:
\begin{eqnarray}
\int_{\rm sh}^r \omega {{\rm d}X\over c^2}
&=&
\omega\int_{\rm sh}^r {\M^2\over 1-\M^2}{{\rm d}r\over v},\\
&\sim&
{\omega r_{\rm sh}\over |v_{\rm sh}|}
{\M_{\rm sh}^2\over \alpha_v+2}\left\lbrack 1-\left({r\over r_{\rm sh}}\right)^{\alpha_v+2}\right\rbrack.
\end{eqnarray} 
With $\omega_r\sim 2\pi |v_{\rm sh}|/r_{\rm sh}$ for a small shock radius this phase shift is not negligible since it  reaches $0.74\sim \pi/4$ at the inner boundary and it is linear in $\omega$.\\
We integrate Eq.~(\ref{def_X}) using Eq.~(\ref{approx_v}) and neglecting $\M^2\ll1$:
\begin{eqnarray}
X_{\rm sh}&\equiv&{r_{\rm sh}v_{\rm sh}\over \alpha_v+1},\\
{X\over X_{\rm sh}}&\sim&\left({r\over r_{\rm sh}}\right)^{\alpha_v+1}.
\end{eqnarray}
The two solutions $Y_0^\pm$ of the homogeneous equation (\ref{homogeneous_Y0}) are approximated as power laws with exponents $\alpha_\pm$. 
\begin{eqnarray}
Y_0^\pm(r)&\equiv&\left({r\over r_{\rm ns}}\right)^{\alpha_\pm}.\label{defYpm}
\end{eqnarray}
Injecting $Y_0^\pm(r)$ into Eq.~(\ref{homogeneous_Y0}),
\begin{eqnarray}
{\alpha_\pm}\left({\alpha_\pm}- \alpha_v-1\right)\left({r\over r_{\rm sh}}\right)^{- 2{\alpha_v}-2}
=
\nonumber\\
-\left\lbrack{\omega^2  r_{\rm sh}^2\over c^2}-\ell(\ell+1) {r_{\rm sh}^2\over r^2}\right\rbrack\left({v_{\rm sh}\over v}\right)^2.
\end{eqnarray}
For $\ell\ge1$, the restoring force in Eq.~(\ref{homogeneous_Y0}) is independent of the frequency
in the region where $\omega\ll\omega_{\rm Lamb}$.
Thus, using Eq.~(\ref{Bersimpl}) for $\ell\ge1$,
\begin{eqnarray}
{\alpha_\pm}\left({\alpha_\pm}- {\alpha_v}-1\right)
&=&\ell(\ell+1)
-\M_{\rm sh}^2
\left({\omega  r_{\rm sh}\over v_{\rm sh}}\right)^2
\left({r\over r_{\rm sh}}\right)^3
,\label{defalphaL}\\
&\sim& \ell(\ell+1)\;\;{\rm if}\;\;\omega\ll\omega_{\rm Lamb}.
\end{eqnarray}
In this region the approximate solution is 
${\alpha_\pm}\equiv\a\pm \b$ with $\a,\b$ defined by Eqs.~(\ref{def_alpha1}) and (\ref{def_alpha2}).
The homogeneous solution $Y_0$ for $\ell\ge1$ is a linear combination of power laws $Y_0^\pm$ satisfying the lower boundary condition (\ref{homogeneous_LBC}):
\begin{eqnarray}
Y_0(r)=\left({r\over r_{\rm ns}}\right)^{\a-\b}+{\cal R}_{\rm ns}\left({r\over r_{\rm ns}}\right)^{\a+\b}\label{defalphaY},\\
{1-\M_{\rm ns}^2\over v_{\rm ns}}\left({\partial Y_0\over\partial r}\right)_{\rm ns}={i\omega \over   c_{\rm ns}^2} Y_0^{\rm ns}.
\end{eqnarray}
With $\M_{\rm ns}\ll1$ and using Eq.~(\ref{approx_M}),
\begin{eqnarray}
\a -\b
+(\a+\b) {\cal R}_{\rm ns}
=
\M_{\rm sh}^2\left({r_{\rm ns}\over r_{\rm sh}}\right)^{2+\alpha_v}
{i\omega r_{\rm sh}\over v_{\rm sh}} 
\left(
 1+{\cal R}_{\rm ns}
\right),\\
{\cal R}_{\rm ns}
=
{
\a-\b
-
\M_{\rm sh}^{4\over\gamma+1}
\left({r_{\rm ns}\over r_{\rm sh}}\right)^{\gamma+3+\alpha_v(3-\gamma)
\over\gamma+1}
{i\omega r_{\rm sh}\over v_{\rm sh}} 
\over
\a+\b
+\M_{\rm sh}^{4\over\gamma+1}
\left({r_{\rm ns}\over r_{\rm sh}}\right)^{\gamma+3+\alpha_v(3-\gamma)\over\gamma+1}
{i\omega r_{\rm sh}\over v_{\rm sh}} 
}.\label{a0_norot}
\end{eqnarray}
From Eq.~(\ref{a0_norot}) we conclude that the coefficient ${\cal R}_{\rm ns}$ is asymptotically independent of $\omega$ when $r_{\rm ns}\ll r_{\rm sh}$.
\begin{eqnarray}
{\cal R}_{\rm ns}
\sim
-{
\a-\b
\over
\a+\b
}
.
\end{eqnarray}
The acoustic function is independent of the frequency in the asymptotic limit $r_{\rm ns}\ll r_{\rm sh}$ for low frequency perturbations $\ell\ge1$ driven by advection such that $\omega r_{\rm sh}/|v_{\rm sh}|\le 1$, as described by Eqs.~(\ref{approx_Y0}) and (\ref{dY0dr}).\\ 
Using the index $0$ to denote the acoustic perturbation associated to the homogeneous solution, we note from Eqs.~(\ref{dvrdA}) with $\delta S=0$ and Eq.~(\ref{def_Y}) that the relation between $\partial Y_0/\partial r$ and $\delta v_r^0$ is 
\begin{eqnarray}
\delta v_r^0&=&
-{1\over \ell(\ell+1)}
{\partial \delta A_0\over\partial r}
,\\
&=&
\left(
{\partial Y_0\over\partial r}
-{i\omega\M^2\over 1-\M^2}{ Y_0\over v}
\right)
{\rm e}^{-\int_{\rm sh} {i\omega\M^2\over 1-\M^2}{{\rm d}r\over v}}.
\end{eqnarray}
The fact that Eq.~(\ref{dY0dr}) imposes $\partial Y_0/\partial r=0$ is approximately compatible with $\delta v_r^0\sim 0$ to the extent that $\M_{\rm ns}\ll1$.\\

\section{Approximation of the eigenfrequency equation for $\gamma=4/3$}

The advection time for $\alpha_v=1$ is approximated using Eq.~(\ref{tadvsph}):
\begin{eqnarray}
{\rm e}^{\int_{\rm sh}^{r}{i\omega\over v_r}{\rm d}r}=
\left({x\over x_{\rm sh}}\right)^{-{i\omega r_{\rm sh}\over |v_{\rm sh}|} }.
\end{eqnarray}
Defining $\delta z\equiv i\omega r_{\rm sh}/ |v_{\rm sh}| +2-\b$ and noting that $\a=1$, the leading terms in Eq.~(\ref{simplest_dispnorot_approx}) for $\ell\ge1$ are :
\begin{eqnarray}
\N(\delta z-2+\b)
+{2\b\M_{\rm sh}^2
\over
\ell(\ell+1)}
x_{\rm sh}^{\delta z-3}
=
\int_{1}^{x_{\rm sh}}
\left({x_{\rm sh}\over x}\right)^{\delta z }
\left(
{1\over x^{2\b}}-1
\right)
{{\rm d}x\over x},\label{leading_disp}
\end{eqnarray}
The integral on the right hand side can be calculated explicitly. If $\delta z\ne0$,
\begin{eqnarray}
\int_{1}^{x_{\rm sh}}
\left({x_{\rm sh}\over x}\right)^{\delta z }
\left(
{1\over x^{2\b}}-1
\right)
{{\rm d}x\over x}
=
{1\over \delta z}\left(
1
-
{ \delta zx_{\rm sh}^{-2\b}
\over 2\b+\delta z}
-
{2\b x_{\rm sh}^{\delta z }
\over 2\b+\delta z}\right),\label{intdw}
\end{eqnarray}
Using Eq.~(\ref{intdw}) in Eq.~(\ref{leading_disp}) we obtain the following equation defining $\delta z$:
\begin{eqnarray}
\left\lbrack
1-
\delta z
\N(\b-2+\delta z)
\right\rbrack
\left(1+{\delta z\over 2\b}\right)
-{\delta z\over 2\b x_{\rm sh}^{2\b}}
=
\nonumber\\
x_{\rm sh}^{\delta z }
\left\lbrack
1
+
\delta z
{2\b\M_{\rm sh}^2
\over
\ell(\ell+1)x_{\rm sh}^{3}}
\left(1+{\delta z\over 2\b}\right)
\right\rbrack.
\label{implicit_analytic_ap}
\end{eqnarray}
Reintroducing the normalized eigenfrequency $Z=i\omega r_{\rm sh}/|v_{\rm sh}|$,
\begin{eqnarray}
{x_{\rm sh}^{\b-2}\over 2\b}
\left\lbrace
\left\lbrack
1-
\left(Z+2-\b\right)
\N\left(Z\right)
\right\rbrack
\left(Z+2+\b\right)
-{Z+2-\b\over x_{\rm sh}^{2\b}}
\right\rbrace
=
\nonumber\\
x_{\rm sh}^{Z}
\left\lbrace
1
+
\left\lbrack
\left(Z+2\right)^2-\b^2
\right\rbrack
{\M_{\rm sh}^2
\over
\ell(\ell+1)x_{\rm sh}^{3}}
\right\rbrace
.
\end{eqnarray}
The approximate advection time $\tau_{\rm adv}^{\rm sh}$ from the shock to the inner boundary, defined by Eq.~(\ref{tadvsph}), is introduced to obtain Eqs.~(\ref{implicit_analytic2}-\ref{implicit_analytic}). 

\end{appendix}

\end{document}